\documentclass[sigconf, nonacm]{acmart}
\usepackage{subcaption}

\usepackage{booktabs}
\usepackage{tabularx}
\usepackage{array}

\AtBeginDocument{%
  }

\setcopyright{acmlicensed}
\copyrightyear{2018}
\acmYear{2018}
\acmDOI{XXXXXXX.XXXXXXX}
\acmConference[Conference acronym 'XX]{Make sure to enter the correct
  conference title from your rights confirmation email}{June 03--05,
  2018}{Woodstock, NY}
\acmISBN{978-1-4503-XXXX-X/2018/06}
\usepackage{listings}
\begin{document}
\raggedbottom

\title{Vibe Interventions: Can Users Personalize and Iterate Their Way to Behavior Change?}


\author{Sally Ahmed}
\email{sallyelf@media.mit.edu}
\affiliation{%
  \institution{MIT Media Lab}
  \city{Cambridge}
  \state{Massachusetts}
  \country{USA}}

\author{Ayse Alomar}
\email{Ayse.Alomar@student.hpi.uni-potsdam.de}
\authornote{Both authors contributed equally to this research.}
\affiliation{%
  \institution{Hasso Plattner Institute}
  \city{Potsdam}
  \country{Germany}}

\author{Tawab Safi}
\email{safi7092@media.mit.edu}
\authornotemark[1]
\affiliation{%
  \institution{MIT Media Lab}
  \city{Cambridge}
  \state{Massachusetts}
  \country{USA}}

\author{Pattie Maes}
\email{pattie@media.mit.edu}
\affiliation{%
  \institution{MIT Media Lab}
  \city{Cambridge}
  \state{Massachusetts}
  \country{USA}}

  \author{Pat Pataranutaporn}
  \email{patpat@media.mit.edu}
\affiliation{%
  \institution{MIT Media Lab}
  \city{Cambridge}
  \state{Massachusetts}
  \country{USA}}

\renewcommand{\shortauthors}{Ahmed et al.}

\renewcommand{\shortauthors}{Ahmed et al.}

\begin{abstract}
Recent advances in large language models allow a working application to be generated from a natural-language description, an approach referred to as generative user interfaces. Their application to behavior change remains largely unexplored, and we provide one of the first controlled investigations of it. We distinguish two components of authorship: iteration, the revision of a generated artifact in natural language, and personalization, the conditioning of generation on stated context. In a pre-registered between-subjects experiment (N = 300, 60 per condition), we crossed these two factors and added a static intervention as a control, yielding five conditions; participants stated a goal and received a generated intervention. The personalized generative condition ranked highest on every self-report outcome, including ownership, satisfaction, and intended adoption. Randomized iteration had no detectable effect, and higher revision counts were negatively associated with evaluative ratings. This work establishes which component of authorship carries motivational value in generative behavior change tools.
\end{abstract}




\keywords{Do, Not, Use, This, Code, Put, the, Correct, Terms, for,
  Your, Paper}
\begin{teaserfigure}
  \centering
  \includegraphics[width=\textwidth]{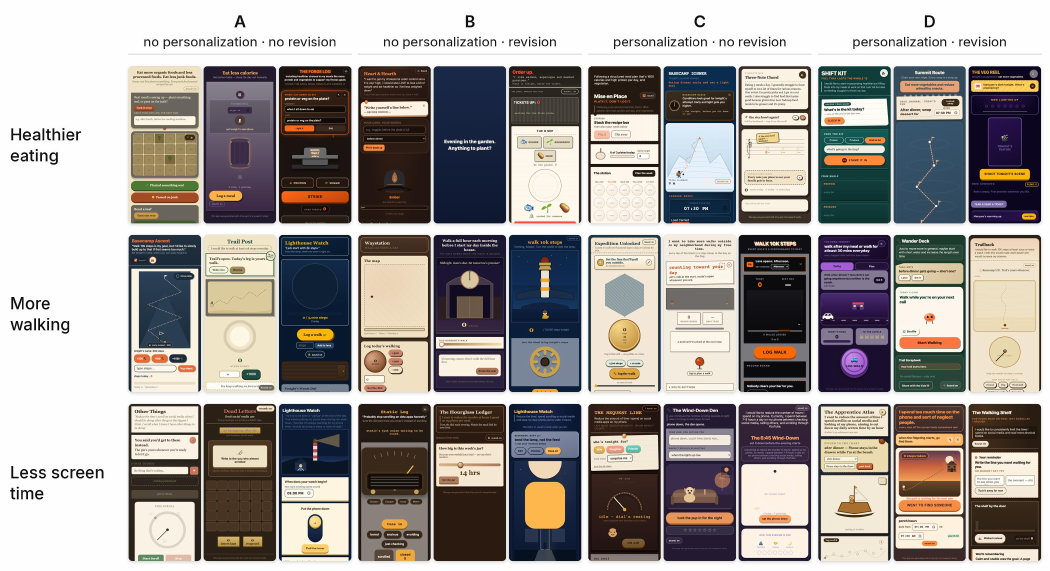}
  \caption{Nine generated apps from each condition, three per goal.
  (A)~GenUI App, (B)~Iterative GenUI App, (C)~Personalized GenUI App,
  (D)~Personalized + Iterative GenUI App. Each app was generated for a
  different participant from their own stated goal.}
  \Description{A grid of 36 smartphone app screenshots arranged in three rows
  by goal and four column groups by condition. Each screenshot shows a
  visually distinct app interface, differing from the others in layout,
  color, typography and framing.}
  \label{fig:teaser}
\end{teaserfigure}

\received{20 February 2007}
\received[revised]{12 March 2009}
\received[accepted]{5 June 2009}

\maketitle

\section{Introduction}
Imagine an app built around you: your goals, your interests, the things that
actually move you, built for the single purpose of getting you to the version of
yourself you want to become. Building software for one person is now something
we can experiment with rather than only describe.

Generative UI (GenUI) is the case where a language model responds to a user's
query by producing an interactive interface \cite{chen2026generative}, written at
run time for that user rather than specified in advance for a population. The
underlying practice is not new. Karpathy coined the term \emph{vibe coding} in 2025
\cite{karpathy2025vibe}, and with tools such as Lovable, Claude Artifacts, and
ChatGPT, describing what you want and letting a model write the code has moved
from a curiosity to a default, available to people who do not write code. What
this makes possible is software for one: an application built by a single person,
for that person.

Precision medicine has pursued a version of this promise for two decades,
matching treatment to a patient's biomarkers and genome, thus tailoring the intervention to the
patient \cite{collins2015precision, ashley2016precision}. Although this offers a fit
designed for the patient, it does not provide them with the autonomy to design this intervention for themselves. Vibe
coding can lift this constraint: the recipient can now specify and revise the
artifact themselves. Applied beyond serious medical conditions to the broader
space of well-being, these two threads suggest an intervention that is both
targeted to and authored by the same person, raising the question of which
half is doing the work.

Behavior change offers a natural setting in which to examine this question, as the generated artifact is authored by and addressed to the same person. Self-Determination Theory predicts that
interventions developed by the subject themselves should provide a greater sense of autonomy compared to interventions they had no prior involvement in designing, and that internalized motivation should outlast imposed control
\cite{ryan2000selfdetermination,alberts2026designing}. If authorship matters
anywhere, it should matter here.

We therefore decompose authorship into two components that are usually bundled
together. Personalization entails tailoring an intervention to an individual's interests and what makes them distinct. Iteration is defined as the capacity of individuals to revise the
generated result toward their intended intervention. Sundar and Marathe conceptually separate
these in their self-as-source and other-as-source framework
\cite{sundar2010personalization}, but existing experimental work on generative
interventions has treated authorship as a single unit, leaving the two factors
uncrossed.

This study asks how personalization and iteration affect
autonomy-related outcomes and stated intentions to act on the proposed behavioral change. Three hundred participants (60 per condition) named a goal in one of three domains: walking more, reducing screen time, or
eating healthier. Following this, they specified their desired persuasive features and received an
intervention generated in their browser during the session. Two factors were
crossed between subjects: personalization, namely whether generation was conditioned on
the participant's stated context, and iteration, specifically whether the participant could
revise the result in natural language. Crossing the two factors yields four generative conditions. These are the GenUI App, which has neither factor, the Iterative GenUI App, the Personalized GenUI App, and the Personalized and Iterative GenUI App. A fifth condition delivered a Static Intervention, a fixed application mimicking existing behavior change apps.

We report the pre-registered outcome family first, followed by a characterization of what participants built, and then the exploratory comparisons. Relative to the GenUI App, the Personalized GenUI App was associated with higher psychological ownership, and participants in that condition also rated the intervention as more personally built, indicating that the personalization manipulation was legible in the system's output. Perceived choice ran in the same direction but did not survive correction for multiple comparisons. Across conditions, the generated interventions converged on a narrow set of components, most frequently reminders, activity feedback, and self-monitoring, and the most common revision request was the addition of a further feature. Within the iterative conditions, participants who revised more reported lower satisfaction, ownership, adoption, and perceived value. No pre-registered outcome showed an effect of generative condition relative to the Static Intervention. Participants valued personalization they could recognize, citing it among their reasons for willingness to adopt the intervention, and reported difficulty in translating their intentions for revision into prompts.

We make four contributions:
\begin{enumerate}

    \item \textbf{Empirical study.} We provide one of the first controlled
    investigations of how the components of generative user interfaces,
    personalization and iteration, affect motivational outcomes in
    behavior change. Prior work has explored whether generated interventions
    outperform non-generated ones. Here, the question is reframed as which
    component of authorship is actually effective, factorially separating two
    components that are ordinarily bundled.

    \item \textbf{Insights on iteration.} We offer insights into how iterative interactions with GenUI can be more effectively designed, including counterintuitive findings regarding how iteration may negatively influence user outcomes.
    
    \item \textbf{User insights.} We characterize how users seek to engage with GenUI for behavior change, including the types of features they prioritize, how they revise their systems, and their preferences for interacting with these systems.
    

    \item \textbf{Design implications and artifacts.} Although GenUI systems in their current form do not yet fully support effective behavior-change interventions, we identify key design opportunities and provide recommendations for improving their effectiveness and overall performance.
    

\end{enumerate}

\section{Related Work}

This study builds on interdisciplinary research from psychology, human-computer interaction, and artificial intelligence. In particular, we extend knowledge from HCI research on digital behavior change and generative interfaces.

\subsection{Interfaces for Behavior Change}

Many studies have used behavior change theories and techniques such as goal-setting, self-monitoring, and feedback through an app-based interface to support users in pursuing their goals \cite{mair2023effective,zhu2024digital,michie2013taxonomy}. To connect app design with motivational theory, Villalobos-Zúñiga and Cherubini analyzed 208 behavior-change applications and identified 12 design features, which they classified according to the basic psychological needs of autonomy, competence, and relatedness proposed by Self-Determination Theory~\cite{villaloboszuniga2020apps}. Their taxonomy provides a framework for considering how app features may support motivation and informed the persuasive features available in the present study. 

Recent work on Self-Determination Theory argues that behavior-change technologies should support autonomy, competence, and relatedness in ways that cultivate self-determined motivation \cite{alberts2026designing,peters2018designing,villaloboszuniga2020apps}. This also applies to digital wellbeing, where HCI research examines how feedback, goal setting, blocking, and interface friction can support attention and screen-time self-regulation while preserving autonomy \cite{hiniker2016mytime,lyngs2019selfcontrol,lukoff2018meaningful}. 

The next wave of research on digital interventions can use LLMs to expand this space and create interventions based on what works best for users and their needs, not just by chatting with a bot, but also by generating the interventions themselves and embedding users' context and goals into them \cite{joerke2025gptcoach,joerke2026bloom}.  Generative interfaces are described as systems in which LLMs proactively construct task-specific user interfaces and iteratively edit and tweak them in response to user input, a significant departure from the linear request–response structure of the conversational interfaces people have used in recent years \cite{chen2026generative,cao2025generative,wu2024uicoder}. LLM-generated interfaces are part of a long-standing HCI tradition of end-user development, which seeks to enable non-programmers to create or modify software for their own needs \cite{lieberman2006enduser,ko2011state,fischer2004metadesign,litt2022potluck,litt2025malleable}. Currently, these interfaces might provide conversational guidance, create plans to change behavior, or change how the interaction is structured \cite{joerke2025gptcoach,chen2026generative}.

HCI research has explored how LLMs can support people's physical activity choices, for example through a chatbot that uses motivational interviewing and the Active Choices program \cite{joerke2025gptcoach}. In this study, the system used both users’ chats with the agent and their wearable data to create personalized physical activity plans \cite{joerke2025gptcoach}. Bloom subsequently integrated an LLM coach with an interface through which users can interact with and refine weekly plans, progress visualizations, reminders, and an ambient garden display \cite{joerke2026bloom}. The results showed an impact on various psychological and engagement outcomes, yet not on participants' physical activity \cite{joerke2026bloom}.

Related work has generated intervention experiences immediately before use; one such system uses the participant's context and chooses the best intervention for them by generating CBT-informed candidate activities \cite{bhattacharjee2026generative}. This system led to immediate stress reduction and a better reported user experience than an LLM-based cognitive-restructuring control \cite{bhattacharjee2026generative}. These findings demonstrate a shift toward more personalized conversations with agents, which can evolve into applications that use LLMs to change behavior and even create interventions on the fly \cite{joerke2026bloom,joerke2025gptcoach,bhattacharjee2026generative}. Researchers have studied how the generative UI system works and what it can do, but more work is needed to establish which psychological effects personalization produces.

\subsection{Personalization for Behavior Change}

Design research has identified personalization as the adaptation of aspects of an information system to a user category \cite{fan2006personalization}. Self-determination theory does not treat personalization as inherently motivating; rather, personalization can support motivation by satisfying users’ psychological needs \cite{ryan2000selfdetermination,peters2018designing,villaloboszuniga2020apps}. Tailoring, in turn, encompasses strategies such as personalizing messages, offering individual feedback, and aligning content with traits related to the targeted behavior \cite{hawkins2008tailoring}. Research has found that tailored health messages and web-based interventions are more effective than non-tailored ones in producing behavior change, although effects vary across populations, behaviors, and approaches to tailoring \cite{noar2007tailoring,lustria2013meta, opdenakker2014tailoring,kaptein2015personalizing,nahumshani2018jitai}. Gosetto et al. tested preferences for 15 behavior-change and gamification mechanisms and found some preferences to vary with Big Five personality traits \cite{gosetto2025preference,gosetto2026personalizing}. Related HCI research also indicates that people can respond differently to the same persuasive strategies, which motivates the personalization of behavior-change systems \cite{orji2017personalization}. Self-congruity is another potential mechanism: when people believe a product or experience aligns with their ideal or real self-concept, they are more likely to evaluate it positively \cite{sirgy1997selfimage}. An intervention can offer behaviorally appropriate support while remaining detached from a person's self-perception or desire for change. Therefore, whether personalizing the intervention interface itself can improve self-congruence, satisfaction, and willingness to adopt the intervention is still up for debate.

\subsection{Authorship, Ownership, and Autonomy}

He et al.'s research found that judgments of authorship credit varied according to the type, amount, and initiative of a contributor's input \cite{he2025contributions}. People may claim authorship of AI-generated text without feeling that it is their own, meaning authorship and psychological ownership can be separate \cite{draxler2024ghostwriter}. Regardless of legal ownership, psychological ownership is defined as the subjective experience that an outcome is “mine” and can develop through perceived control, intimate knowledge, and self-investment \cite{pierce2001psychological,pierce2003state}. Related evidence from the ``I designed it myself'' effect shows that participating in creation can increase how highly people value an outcome, partly through accomplishment and perceived influence~\cite{franke2010designed}. 

Autonomy is defined as the experience that one’s actions are volitional and self‑endorsed, rather than controlled by external forces~\cite{ryan2000selfdetermination}. Recent HCI research applied this perspective to explore how digital systems might support or undermine motivation and well-being \cite{peters2018designing}. By contrast, sense of agency refers to the feeling of having control over actions \cite{moore2016agency}. In generative systems, agency may depend on whether users can see that their input meaningfully changes the generated outcome \cite{draxler2024ghostwriter,xu2024whatmakesitmine}.

Generative AI further complicates these relationships since the system can generate large parts of an outcome. In human-AI co-creation research, involvement, perceived contribution, control, and the agency attributed to the AI can influence psychological ownership \cite{xu2024whatmakesitmine}. AI-assisted writing is associated with lower psychological ownership compared to independent writing, while more substantial or granular user input can increase ownership \cite{joshi2025writing}. Draxler et al. also found that increasing influence over generated text increased ownership, but personalization alone did not \cite{draxler2024ghostwriter}. Existing systems often combine personalized generation and iterative refinement, leaving their distinct effects on users' agency, ownership, and evaluations unclear.

\section{Methods}
In a pre-registered, between-subjects randomized experiment, 300 participants were recruited through Prolific. Participants completed a pre-session survey about demographics, technology use, goal, and baseline behavioral intention, and then interacted with a web application. Within the application, participants were asked about their preferences and the goal they wanted to work on, and chose the persuasive features for their intervention. Four of the five conditions generated an intervention with a large language model, crossing whether the participant's interest inventory was passed to the model (\textit{personalization}) with whether the participant could revise the result in natural language (\textit{iteration}). The fifth was a static, non-generative intervention matched on behavior change technique coverage. All outcomes were measured at the end of the session. The study was pre-registered on OSF prior to data collection and reviewed by our institutional review board. The following section details the research questions, the design and its rationale, the technical implementation, the protocol, the measures, and the analysis plan.

\subsection{Research Questions}
The study asks whether authorship of a behavior change intervention produces better proximal psychological and behavioral outcomes than receiving one, and whether the two components of authorship separate.
\begin{itemize}
\item[\textbf{RQ1a.}] Does personalization (passing a participant's interests to the generating model) make them more inclined to change their behavior than generating from their goal alone?

\item[\textbf{RQ1b.}] Does iteration (allowing a participant to revise the generated intervention in natural language) make them more inclined to change their behavior than receiving a fixed generation?

\item[\textbf{RQ2.}] Which component of authorship does more for congruency, ownership, agency, and commitment: iteration or personalization?

\item[\textbf{RQ3.}] What do participants build when given authorship over a behavior change intervention, and what do they iterate toward and away from during iteration?
\end{itemize}

\subsection{Design}
The study uses a five-condition between-subjects design: a static, non-generative control (Static Intervention App) plus a 2×2 factorial crossing \textit{personalization} (the participant's interest inventory is passed to the generating model vs. withheld) with \textit{iteration} (revision of the generated intervention is permitted vs. not), with the GenUI App acting as the generative control.


\subsubsection{Conditions}
\begin{table}[t]
  \caption{Experimental design. Five conditions: a static, non-generative control (Static Intervention App) plus a $2\times2$ factorial crossing personalization with iteration.}
  \Description{Table listing the five study conditions with their personalization and iteration settings and the number of participants per condition.}
  \label{tab:design}
  \resizebox{\columnwidth}{!}{%
  \begin{tabular}{lccr}
    \toprule
    \textbf{Condition} & \textbf{Personalization} & \textbf{Iteration} & \textbf{\textit{n}} \\
    \midrule
    Static Intervention App& --- (static control) & --- & 60\\
    GenUI App& No  & No  & 60\\
    Iterative GenUI App& No  & Yes & 60\\
    Personalized GenUI App& Yes & No  & 60\\
    Personalized + Iterative GenUI App& Yes & Yes & 60\\
    \midrule
    \textbf{Total} & & & \textbf{300} \\
    \bottomrule
  \end{tabular}}
\end{table}

Each condition is load-bearing. Static Intervention App is the only condition with no model involvement: a generic tracker that holds goal-anchored functionality constant while removing generation, and so anchors any claim about authorship. GenUI App is generative but withholds both factors, producing an intervention from the participant's goal and selected persuasive features alone; it is the reference cell against which personalization and iteration are each measured. Iterative GenUI App and Personalized GenUI App are the single-factor cells and constitute the study's central comparison, setting iteration against personalization directly. Personalized and Iterative GenUI App is the only cell in which the two factors co-occur and is therefore the sole source of information about whether their effects are additive.

\subsection{Study Design Rationale}



\subsubsection{Why decompose authorship}
Generative authorship is already in users' hands. Tools built on the premise of generative authorship, including Lovable and Claude Artifacts, now let people assemble working software for an audience of one. Behavior change interventions are a natural target, since the thing being designed is a system for changing the designer's own conduct. What these tools deliver as authorship bundles two things: the system can build around what it knows about a person, and the person can revise what it builds. Psychological ownership theory treats intimate knowledge and perceived control as separate routes to feeling that an outcome is one's own \cite{pierce2001psychological}. In this study we cross these two aspects factorially, with a generative condition receiving neither as a reference.

\subsubsection{Persuasive features for users to choose}
Our set of available features draws on the taxonomy of Villalobos-Zúñiga and Cherubini~\cite{villaloboszuniga2020apps}, who analyzed 208 behavior change applications and identified 12 design features mapped onto the basic psychological needs posited by Self-Determination Theory. We made seven available to participants: reminders, motivational messages, pre-commitment, activity feedback, history, self-monitoring, and messaging a peer.

Of the five features we did not offer, goal setting was fixed by the study design, since participants worked within an assigned behavior domain rather than selecting a goal themselves. The remaining exclusions were driven by what a browser-based task completed in under 30 minutes can represent without fabrication. Messaging a peer is an outbound action that a participant can fully specify from their own side. Performance sharing, peer comparison, and challenging a peer all require inbound content from another user, so representing them would have meant inventing peer identities and performance data.

\subsubsection{Why walking, reducing screen time, and eating healthier}

All three are among the changes people most commonly
report wanting to make of their own accord. When asked to formulate goals
freely, participants in a large general-population study most often named
physical health, weight, and eating habits \cite{oscarsson2020large}, and
self-trackers most frequently monitor activity and food
\cite{choe2014understanding}. Screen time reduction is now comparably
widespread: 53\% of U.S. adults report spending too much time on their
smartphone, and 45\% attempted to cut back in the past year, yet only a
quarter of those who tried judged themselves very or extremely successful
\cite{pew2026smartphone}. 


\subsubsection{Static Intervention App as a non-generative control}
The control is a plain dashboard in line with current health trackers such as Apple Health. It holds goal-anchored functionality constant while removing generation entirely, so any difference between it and the generative conditions is attributable to generation rather than to what the intervention offers. 

\subsubsection{Risk of Self-Authored Interventions}
Self-authorship shifts control over goal content to the person whose behavior is at stake. When difficulty is held constant, self-set and assigned goals perform comparably \cite{latham1988resolving}; difficulty, not origin, carries the effect \cite{locke2002building}. The risk is that a person setting a goal without an external anchor may set an easier one. Our interface bars numeric targets, but difficulty remains uncontrolled in the free-text goal statements, which we code post hoc.

The self as designer may also be a poor forecaster of the self as adherent. Immediate rewards predict persistence while delayed rewards do not \cite{woolley2017immediate}, and people undervalue intrinsic incentives when evaluating an activity from outside it \cite{woolley2015experience}. A participant designing an intervention is by construction outside it, and so optimizes against a model of the future self that underweights what would sustain use.

Finally, safety is a property of the generating system, not of the authorship arrangement. Our prompts and filters constrain output for goals bordering on disordered eating or self-monitoring; end-user tooling without comparable guardrails would carry the same affordances without the same limits~\cite{dechoudhury2023benefits, moore2025expressing}. 

\subsection{Technical Implementation}
\label{sec:technical}

\subsubsection{Generation pipeline.}
Each intervention is produced by two agents, both Claude Sonnet~5, chosen over Opus~5 for latency; Discovery is a scripted survey and involves no model. The \emph{ideation} agent receives the goal category, the stated goal, and the guidance for each selected behavior change technique (BCT), and returns a structured build plan: the app's concept, mechanic and visual language, how each technique appears as a working surface, and the actions the app records. The \emph{writing} agent turns that plan into one self-contained HTML document with inline CSS, JavaScript, and hand-drawn graphics. The app is then checked: that it is a complete, self-contained document, that every selected technique and the AI disclosure are present, and that it uses no failure or body-weight language; an app that fails is returned with its failing checks to a repair agent (Appendix~\ref{app:prompts}). Only an app that passes, and renders without error in the participant's browser, is shown. Of 240 first builds, 21 needed one repair and none needed a second; ideation took a median of 37\,s and writing 200\,s. The prompts were frozen before enrollment and identical for every participant, and every call is stored with its rendered prompt, raw output, validation report, and token counts.
 
\subsubsection{Where personalization and iteration enter.}
Personalization is a difference in prompt content and nothing else: in the personalization conditions the ideation and writing prompts include the participant's three inventory answers verbatim and a mandate to build for this person and nobody else; in the other conditions they carry a generic mandate, and the goal, stated goal, and technique guidance are identical. Iteration enters only after the first build. A revision re-runs the writing agent with the same plan plus the participant's requested change, without the current document or earlier requests, so each revision regenerates the whole app inside a fixed plan; of 257 revisions, 27 needed a repair. The cap of three revisions is enforced server-side.

\subsubsection{Safety checks.}
Generated interventions can nudge with the wrong motives as easily as the right ones, and the pathways are documented: numeric targets, weight and body framing, and failure language are how diet and activity content reinforces restriction and self-criticism~\cite{winecoff2026symptoms, dechoudhury2023benefits, moore2025expressing}. Following Bloom~\cite{joerke2026bloom} and prior generative interventions~\cite{bhattacharjee2026generative}, we fixed a set of constraints in advance and enforced them at three points. At Discovery, a stated goal naming a diagnosis, a medication, or a weight, BMI, or calorie target is declined and a non-clinical goal requested. In generation, the prompts forbid any numeric target the participant did not state; weight, body-measurement, or before-and-after content; comparison against other people or norms; failure states such as breakable streaks or shame language; invented facts about the participant; and quotations attributed to real people that the model cannot vouch for. After generation, an app is rejected if its source contains any term from a fixed list of failure and body-related language, omits the AI disclosure, or loads an external resource, and the offending matches are passed to the repair agent. These checks are lexical and prompt-level rather than clinical review. Because prompts carry participants' own words, every model call was made with zero data retention, so the model provider kept no prompts or outputs, and requests identified participants only by a hashed ID.

\subsection{Experimental Procedure}

\begin{figure*}[t]
  \centering
  \includegraphics[width=0.75\textwidth]{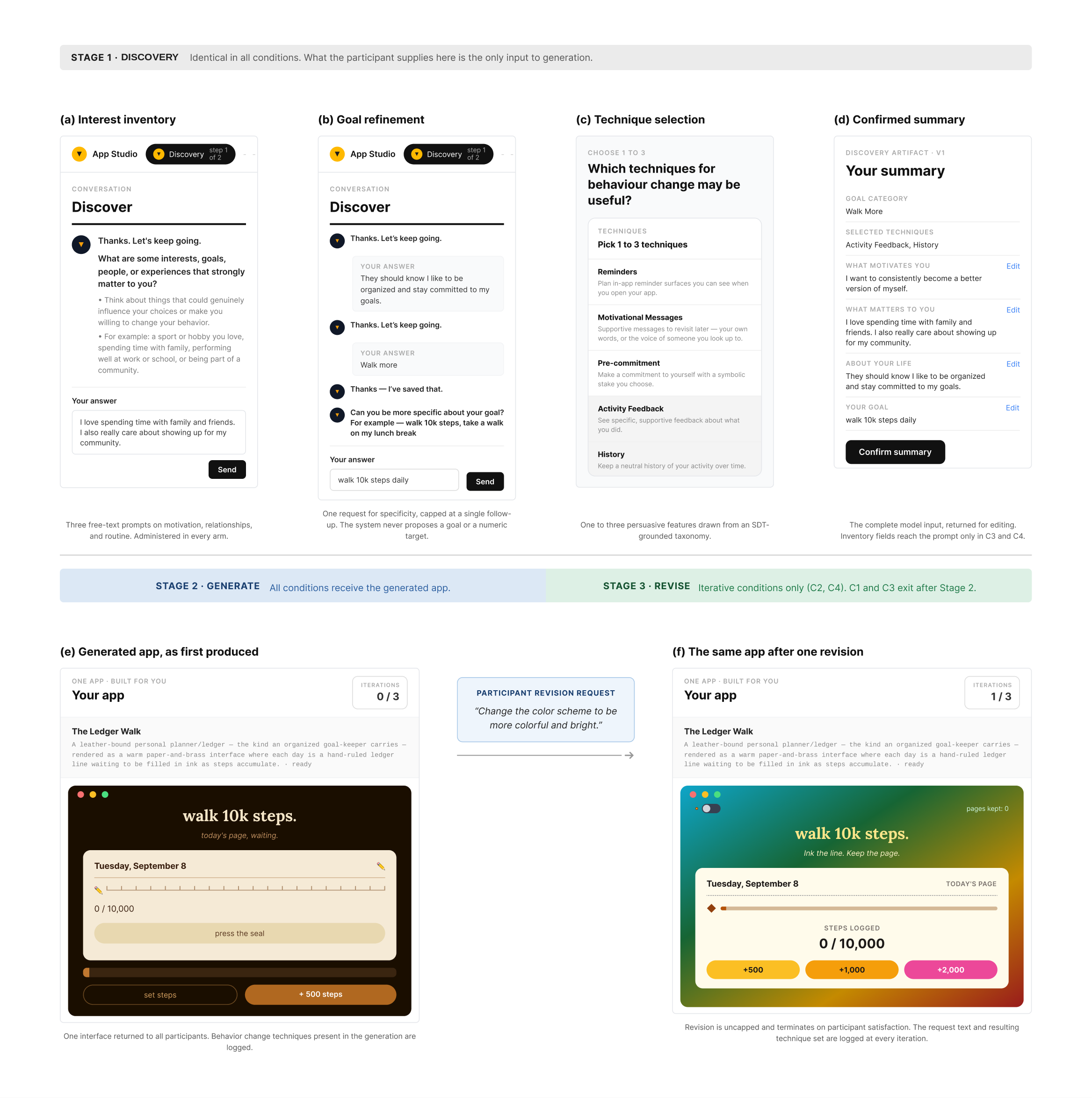}
  \caption{The three-stage study pipeline. \textbf{Stage 1 (Discovery)} was identical
  across conditions and produced the only input to generation: (a) three free-text prompts on motivation, relationships, and routine; (b) a goal statement with at most one request for specificity, where the system never proposed a goal or a numeric target; (c) selection of one to three persuasive techniques from an SDT-grounded taxonomy; and (d) an editable summary of the complete model input. Inventory fields}
  \label{fig:myimage}
\end{figure*}

\subsubsection{Pre-session survey}

After consent, participants reported demographics (age, gender, country, employment), current use of activity trackers or wellness applications, and prior experience using generative AI to create or customize an interface. Participants also selected a goal and answered a three-item baseline goal-anchored behavioral intention scale adapted from Theory of Planned Behavior measurement guidance~\cite{ajzen1991,ajzen2006}.

\subsubsection{Discovery}
After entering the application, participants first complete a short interest inventory of three open-ended questions: what tends to make them feel motivated or excited to do something; what interests, goals, or experiences matter strongly to them; and what someone designing for their life would need to know about them. Every participant answers all three regardless of condition. Their responses are passed to the generating model only in the personalization conditions (Personalized GenUI App, Personalized and Iterative GenUI App); administering the inventory universally and gating only its transmission makes the manipulation the flow of information rather than the act of introspection.

Participants then select a goal category -- walk more, eat healthier, or reduce screen time -- and are asked exactly once to make the goal specific, for example, "Can you be more specific about your goal? For example: walk 10k steps, take a walk on my lunch break." Finally, participants select the persuasive features they want their intervention built around, choosing up to three from six: reminders, pre-commitment, motivational messages, activity feedback, log/history, and involving a friend. The set is drawn from an SDT-grounded taxonomy of persuasive application features and spans the three basic psychological needs, with reminders, pre-commitment, and motivational messages supporting autonomy; activity feedback and log/history supporting competence; and involving a friend supporting relatedness~\cite{villaloboszuniga2020apps}. Capping selection at three holds the scope of the generation task roughly constant across participants, so that interventions do not vary in breadth as a function of how many features a participant happened to endorse.

\subsubsection{Implementation}
Participants interact with their intervention. In the iteration conditions (Iterative GenUI App, Personalized and Iterative GenUI App) they may request modifications in natural language, up to three revisions; each revision triggers a fresh generation, and both the request text and the resulting artifact are logged. In the non-iteration conditions (GenUI App, Personalized GenUI App) they interact with the intervention as generated and no revision affordance is presented. Static Intervention App participants interact with the static intervention.

\subsubsection{Post-session Survey}
After interacting with their intervention, users completed the outcome battery described in Section~\ref{sec:measures}, assessing behavioral intention~\cite{ajzen1991,ajzen2006}, actual-self congruency~\cite{sirgy1997selfimage,wu2020}, psychological ownership~\cite{caspi2011}, IMI Perceived Choice and Value/Usefulness~\cite{mcauley1989}, satisfaction~\cite{zhou2019}, adoption likelihood~\cite{venkatesh2000}, mental demand~\cite{hart1988}, and sense of agency (Sense of Agency Scale)~\cite{tapal2017,draxler2024ghostwriter}. Participants also reported whether they would have made changes to the app, and those who answered yes were asked to describe their desired changes in an open-ended item. All participants answered open-ended items asking what they liked about the app and its creation process, and why they would or would not use it in everyday life. Finally, participants indicated whether they would pledge to work toward their selected goal over the following week, reported whether the app had functioned sufficiently to complete the study in a free-text item, and could provide additional feedback in another open-ended item. Full survey items are detailed in Appendix~\ref{app:post_survey}.

\subsection{Participants}

\subsubsection{Recruitment and assignment}
Participants were recruited through a Prolific listing and randomized to one of the five conditions on entering the application; assignment was made application-side and was not visible in the study listing. Inclusion criteria
were: 18 years or older, access to a laptop or desktop computer and a supported browser, and a self-reported goal in one of the study's three goal categories. Condition-specific compensation was disclosed only after assignment, so
participants could not select into a condition on the basis of pay. Payment was set in proportion to each condition's expected duration---\$8 (Static Intervention App), \$10 (GenUI App), \$12 (Personalized GenUI App), and \$14 (Iterative GenUI App
and Personalized and Iterative GenUI App)---targeting an effective rate of approximately \$30 per hour in the generative conditions at the median session span.
The static control cleared a higher effective rate, as its median span was shorter than the payment schedule anticipated.

\subsubsection{Sample characteristics}
Table~\ref{tab:demographics} reports the analyzed sample by condition. Participants
were 18--45 years old ($M = 32.4$, $SD = 7.0$); two participants entered
implausible ages and are excluded from the age statistics only. The sample was
close to evenly split by sex, predominantly employed full-time (64.0\%), and
56.3\% reported prior experience using a generative AI tool to make an app,
website, or interface. Goal selection clustered on physical health: 119
participants (39.7\%) chose healthier eating, 115 (38.3\%) increasing daily
steps, and 66 (22.0\%) reducing screen time. Baseline goal-anchored behavioral
intention was comparable across conditions ($M = 5.48$--$5.78$ on a 7-point scale),
as were baseline motivation and goal importance.

Median study span, measured from pre-session survey start to post-session
submission, scaled with condition complexity: 8.4~min (Static Intervention App),
19.8~min (GenUI App), 21.2~min (Personalized GenUI App), 28.4~min (Iterative
GenUI App), and 32.4~min (Personalized and Iterative GenUI App). Distributions
were right-skewed, so we report medians. Seven participants completed the study
in under five minutes, all of them in the static control; they are retained in
all analyses and flagged in the released data.

\begin{table*}[t]
\caption{Analysed sample by condition ($n = 60$ per condition, $N = 300$).}
\label{tab:demographics}
\small
\begin{tabular}{lrrrrrr}
\toprule
& C0 & C1 & C2 & C3 & C4 & Total \\
\midrule
\textit{Age} $M$ (SD)\textsuperscript{a}
 & 31.1 (7.5) & 32.2 (7.3) & 32.5 (7.3) & 32.8 (6.9) & 33.6 (6.2) & 32.4 (7.0) \\
\addlinespace
\textit{Gender} & & & & & & \\
\quad Female & 29 & 33 & 25 & 28 & 29 & 144 \\
\quad Male & 29 & 25 & 33 & 31 & 31 & 149 \\
\quad Non-binary & 1 & 2 & 1 & 0 & 0 & 4 \\
\quad Prefer not to say & 1 & 0 & 1 & 1 & 0 & 3 \\
\addlinespace
\textit{Employment} & & & & & & \\
\quad Full-time & 31 & 40 & 37 & 41 & 43 & 192 \\
\quad Part-time & 15 & 8 & 7 & 11 & 6 & 47 \\
\quad Unemployed & 12 & 10 & 12 & 3 & 10 & 47 \\
\quad Student & 2 & 2 & 4 & 5 & 1 & 14 \\
\addlinespace
\textit{Prior generative-AI interface use} & & & & & & \\
\quad Yes & 32 & 33 & 25 & 40 & 39 & 169 \\
\addlinespace
\textit{Goal selected} & & & & & & \\
\quad Healthier eating & 25 & 14 & 29 & 22 & 29 & 119 \\
\quad Daily steps / walking & 20 & 27 & 22 & 26 & 20 & 115 \\
\quad Reduce screen time & 15 & 19 & 9 & 12 & 11 & 66 \\
\addlinespace
\textit{Prior goal-support app use} & & & & & & \\
\quad Currently uses one & 12 & 13 & 15 & 15 & 16 & 71 \\
\quad Used one in the past & 22 & 27 & 23 & 23 & 26 & 121 \\
\quad Never used one & 26 & 20 & 22 & 22 & 18 & 108 \\
\addlinespace
Baseline intention $M$ (SD)
 & 5.78 (1.02) & 5.48 (1.17) & 5.57 (1.48) & 5.74 (1.10) & 5.52 (1.18) & 5.62 (1.20) \\
Median study span (min)
 & 8.4 & 19.8 & 28.4 & 21.2 & 32.4 & 22.5 \\
\bottomrule
\multicolumn{7}{l}{\footnotesize \textsuperscript{a}Two participants (one each in
C2 and C3) entered implausible ages and are excluded from age statistics only.}
\end{tabular}
\end{table*}

\subsubsection{Scale reliability}
Table~\ref{tab:reliability} reports internal consistency for every multi-item
scale. Reliability was high for behavioral intention, IMI Value/Usefulness,
actual-self congruency, and psychological ownership ($\alpha \geq .92$),
acceptable for perceived personalization ($\alpha = .78$) and the Sense of
Agency Scale ($\alpha = .71$), and \emph{low} for IMI Perceived Choice
($\alpha = .58$, $\omega = .73$).

\begin{table}[t]
\caption{Internal consistency of multi-item scales ($N = 300$).}
\label{tab:reliability}
\small
\setlength{\tabcolsep}{3.5pt}
\begin{tabular}{lrrrrr}
\toprule
Scale & Items & $\alpha$ & $\omega$ & $M$ & $SD$ \\
\midrule
Behavioral intention (baseline) & 3 & .95 & .96 & 5.62 & 1.20 \\
Behavioral intention (post-session) & 3 & .96 & .96 & 5.89 & 1.11 \\
IMI Perceived Choice & 3 & .58 & .73 & 5.40 & 1.16 \\
IMI Value / Usefulness & 3 & .92 & .92 & 5.36 & 1.42 \\
Actual-self congruency & 3 & .96 & .96 & 4.52 & 1.76 \\
Perceived personalization & 3 & .78 & .81 & 4.61 & 1.54 \\
Psychological ownership\textsuperscript{a} & 3 & .95 & .95 & 3.03 & 1.34 \\
Sense of agency & 7 & .71 & .73 & 4.66 & 1.01 \\
\bottomrule
\multicolumn{6}{l}{\footnotesize \textsuperscript{a}1--5 scale; all others 1--7.
$\omega$ estimated from a single-factor principal-axis solution.}
\end{tabular}
\end{table}

\subsection{Measures}
 \subsubsection{Self-report outcomes}
\label{sec:measures}

Goal-anchored behavioral intention was measured with three items following Theory of Planned Behavior measurement 
guidance~\cite{ajzen1991,ajzen2006}, administered at baseline and after the session with identical stems and the participant's own goal piped into each item. Actual-self congruency was measured with three items adapted from self-congruity measurement~\cite{sirgy1997selfimage,wu2020}, capturing whether the intervention matched how participants saw themselves; the actual-self rather than the ideal-self formulation was used, since the construct of interest was fit with who the participant was rather than who they wished to become. Psychological ownership was measured with three items adapted from~\cite{caspi2011}, with one item explicitly acknowledging the model's contribution so that participants were not required to choose between felt ownership and accurate attribution.

Intrinsic motivation was measured with the Perceived Choice and Value/Usefulness subscales of the Intrinsic Motivation Inventory~\cite{mcauley1989}, three items each, adapted so that ``this activity'' referred to creating, reviewing, and selecting the intervention. The subscales were analyzed separately and never composited, as they index autonomy and internalization respectively. Satisfaction and adoption likelihood were each measured with a single item~\cite{zhou2019,venkatesh2000}, the latter followed by an open-ended prompt asking why. Sense of agency was measured with seven items adapted from the Sense of Agency Scale~\cite{tapal2017} and worded for a generative-AI authorship context following~\cite{draxler2024ghostwriter}. Perceived mental demand was measured with a single item adapted from the NASA Task Load Index~\cite{hart1988}. Participants then provided a free-text implementation intention, coded 0--7 for specificity, followed by a binary pledge to work toward their goal over the coming week~\cite{michie2013}. Three open-ended items asked what participants would change, whether they would use the intervention in everyday life and why, and what they liked most.

Manipulation checks were administered for both factors: perceived personalization in all conditions with four items following the computer-tailored health communication literature~\cite{kreuter1999}, checked as the directional contrast between $\{$Personalized GenUI App,\,Personalized and Iterative GenUI App$\}$ and $\{$GenUI App,\,Iterative GenUI App$\}$; and self-reported degree of revision in the iteration conditions only, corroborated against logged iteration counts. Two attention checks were embedded in the post-session battery.

\subsubsection{System-logged measures}
Condition assignment; session timestamps and duration; iteration count and full revision text; free-text input length; behavior change techniques used in each generation; the fully resolved prompt per generation; persuasive features present in the delivered intervention; and screenshots of each generated interface.

\subsection{Statistical Analysis}

Analyses are reported in two tiers. Confirmatory analyses are those specified in the pre-registration; everything else is labeled exploratory and interpreted
descriptively. We adopt this split rather than reporting a single undifferentiated set of contrasts because the study yields more comparisons than it can support inferentially, and the distinction determines which results carry evidential weight.

\subsubsection{Confirmatory analyses}
The pre-registered question was whether generated interventions outperform a competent static one. The confirmatory analysis therefore contrasts the Static Intervention App against the generative conditions on the pre-registered outcome family: behavioral intention, pledge, IMI Perceived Choice, IMI Value/Usefulness, actual-self congruency, and satisfaction. Generative conditions are pooled by whether revision was available, giving two contrasts per outcome (control vs.\ non-iterative conditions, control vs.\ iterative conditions).

Each continuous outcome was modeled with a linear model of the form
$Y = \beta_0 + \beta_1 \text{condition} + \varepsilon$, and post-session behavioral
intention additionally with an ANCOVA adjusting for baseline intention,
$Y_{\text{post}} = \beta_0 + \beta_1 \text{condition} + \beta_2 Y_{\text{pre}} +
\varepsilon$. Baseline adjustment is used only where a genuine pre-measure exists; the remaining outcomes were measured once, post-session, and so are modeled without a covariate. The pledge is binary and was modeled with logistic regression. Familywise error across the six pre-registered outcomes was controlled with the Bonferroni correction; both corrected and uncorrected $p$-values are reported.

Because the study is powered to detect $d = 0.47$--$0.60$ across this family, we
report the minimum detectable effect alongside each null so that the nulls
function as bounds rather than as absence of evidence. We do not interpret any
null as evidence of equivalence.

\subsubsection{Manipulation checks}
Two manipulation checks were pre-specified. Perceived personalization was
measured in all conditions and tested as the directional contrast between the
personalization conditions \{Personalized GenUI App, Personalized and Iterative GenUI
App\} and the non-personalization conditions \{GenUI App, Iterative GenUI App\}.
Self-reported degree of revision was collected in the iteration conditions and
corroborated against the logged revision count.

\subsubsection{Exploratory analyses}
Four sets of analyses are exploratory. First, the decomposition of the
generative conditions: each generative condition contrasted against the GenUI App, and the
$2 \times 2$ factorial among the four generative conditions modelling personalization,
iteration, and their interaction with the GenUI App as the reference cell. These
contrasts speak to which component of authorship carries any effect, but were
not the pre-registered comparison, and outcomes outside the pre-registered
family (ownership, sense of agency, perceived personalization, adoption
likelihood, mental demand) are reported with uncorrected $p$-values and should
be read as descriptive.

Second, the revision-dose analysis within the iteration conditions ($n = 120$),
regressing each outcome on the logged number of revisions requested, adjusting
for condition, baseline behavioral intention, and goal category. Revision count is
participant-chosen rather than assigned, so this analysis is observational and
supports no causal reading.

Third, the LLM-based scoring of free-text goals, self-disclosures, and revision
requests against a fixed rubric, and the association of judge-rated goal
concreteness with self-reported outcomes. Because goal concreteness is a
property of what participants wrote rather than a randomized factor, this
association is observational. We report these results in the appendix. Fourth, thematic analysis of the open-ended responses.

All exploratory tests are reported without familywise correction and are
described as patterns rather than findings. Where an exploratory analysis
produces the same number of significant results as chance would predict, we say
so.

\subsection{Ethics, Pre-registration, and Approval}
This research was reviewed and approved by the Institutional Review Board, protocol number \#XXXXXXXXXXX, and determined exempt. The design and analysis plan were pre-registered on OSF ([Anonymized]) prior to data collection. Participants provide informed consent before any data collection and may withdraw at any point without penalty. No deception is used. Generated interventions are constrained never to produce weight, BMI, body measurements, before/after framing, or comparisons against population norms.

\section{Results}

We report the pre-registered outcome family first, then what participants built, then the exploratory comparisons. Relative to the GenUI App, the Personalized GenUI App was associated with higher psychological ownership; participants in that condition also rated the intervention as more personally built, indicating that the personalization manipulation was legible in the system's output. Perceived choice ran in the same direction but did not survive correction for multiple comparisons. Across conditions, the generated interventions converged on a narrow set of components, most often reminders, activity feedback, and self-monitoring, and the most common revision request was the addition of a further feature. Within the iterative conditions, participants who revised more reported lower satisfaction, ownership, adoption, and perceived value; revision count was participant-chosen, so this association is observational. No generative condition outperformed the Static Intervention on any pre-registered outcome, and none was outperformed by it.

\subsection{Personalized GenUI App is associated with higher ownership and perceived choice}

\begin{figure*}[t]
  \centering
  \begin{subfigure}{\textwidth}
    \centering
    \includegraphics[width=\textwidth]{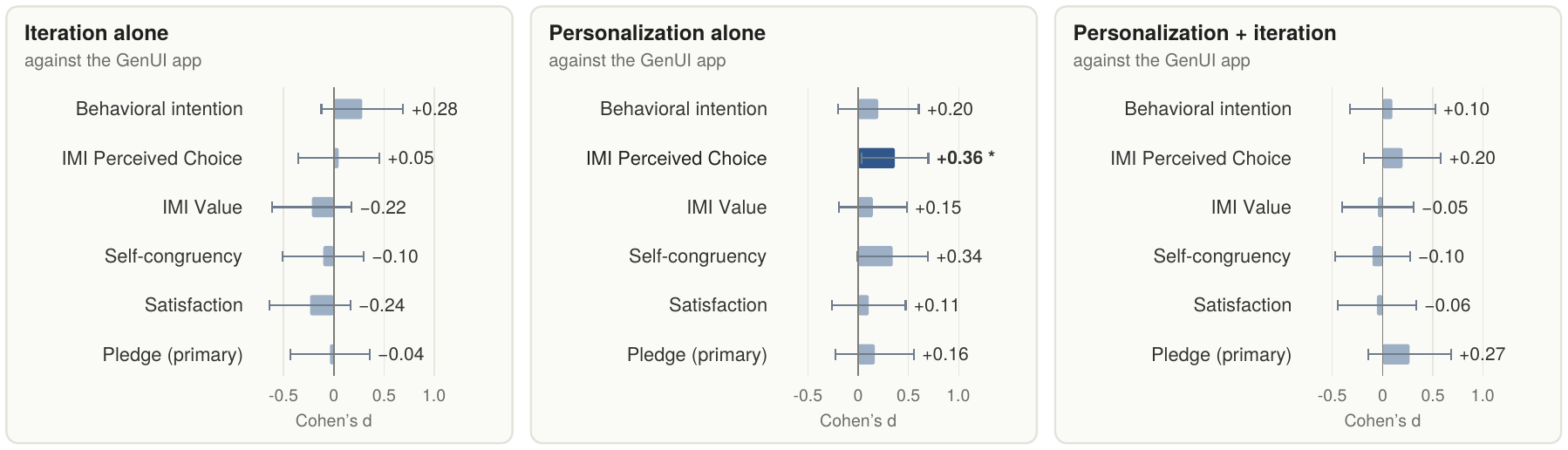}
    \caption{Pre-registered outcome family.}
    \label{fig:d-prereg}
  \end{subfigure}
  \\[6pt]
  \begin{subfigure}{\textwidth}
    \centering
    \includegraphics[width=\textwidth]{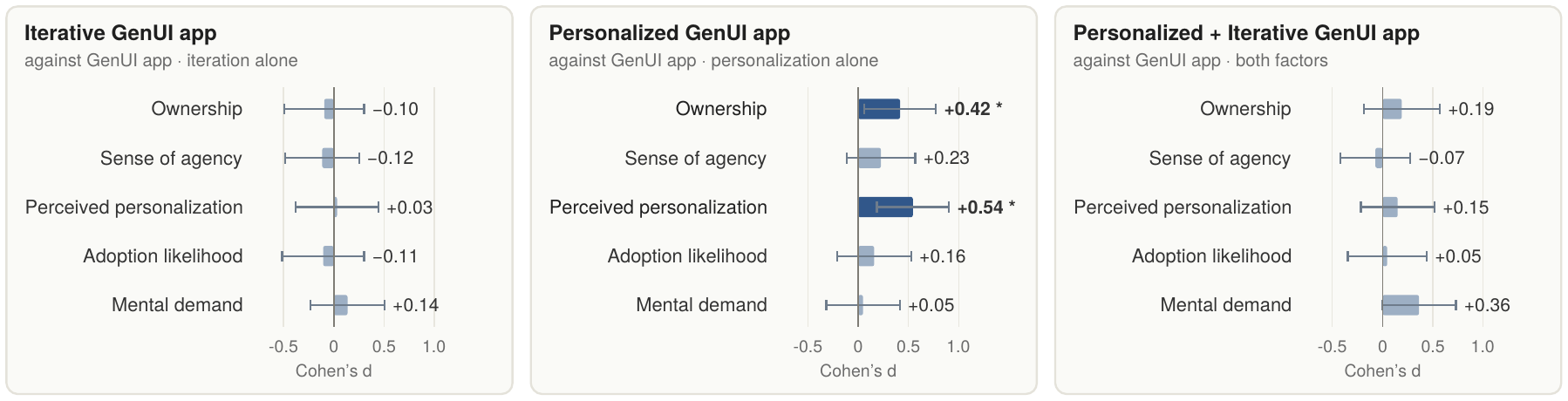}
    \caption{Outcomes outside the pre-registered family. Asterisks mark uncorrected $p < .05$.}
    \label{fig:d-beyond}
  \end{subfigure}
  \caption{Standardized differences between each generative condition and the GenUI App. Bars are Cohen's $d$ with 95\% confidence intervals.}
  \Description{Two stacked panels of three columns each...}
  \label{fig:d-effects}
\end{figure*}

To ask what each factor contributed on its own, we compared each generative
condition against the GenUI App, the condition with neither personalization nor iteration. We report these
descriptively. As Figure~\ref{fig:d-prereg} shows, every confidence interval in
the pre-registered outcome family falls inside the band of effects this study
was powered to detect, and no outcome in that family survives the
pre-registered Holm correction. What follows is therefore a description of
where the conditions differed, not a confirmatory test of why.

The two factors behaved very differently. Adding iteration alone moved nothing:
across the eleven outcomes the largest standardized difference was on
behavioral intention ($d = 0.28$, 95\% CI $[-0.12, 0.69]$), every interval
covered zero, and six of the eleven point estimates were negative. Adding both
factors together was similarly flat, with the exception of mental demand
($d = 0.36$, $[-0.00, 0.73]$); this was the most effortful condition in the study.

Adding personalization alone was the only manipulation associated with
movement, and it moved two outcomes in particular. Psychological ownership was
higher in the Personalized GenUI App than in the GenUI App ($d = 0.42$,
$[0.06, 0.77]$, $p = .022$), as was IMI Perceived Choice
($d = 0.36$, $[0.03, 0.70]$, $p = .032$). Actual-self congruency sat
just at the boundary ($d = 0.34$, $[-0.01, 0.69]$, $p = .053$). The
manipulation check moved furthest of anything measured: participants in the
personalized condition rated the intervention as more personally built ($d = 0.54$,
$[0.19, 0.90]$, $p = .003$), confirming that the interest inventory reached the
model and was legible in its output.

\subsection{What participants chose and built}

\begin{figure*}[t]
  \centering
  \includegraphics[width=0.8\textwidth]{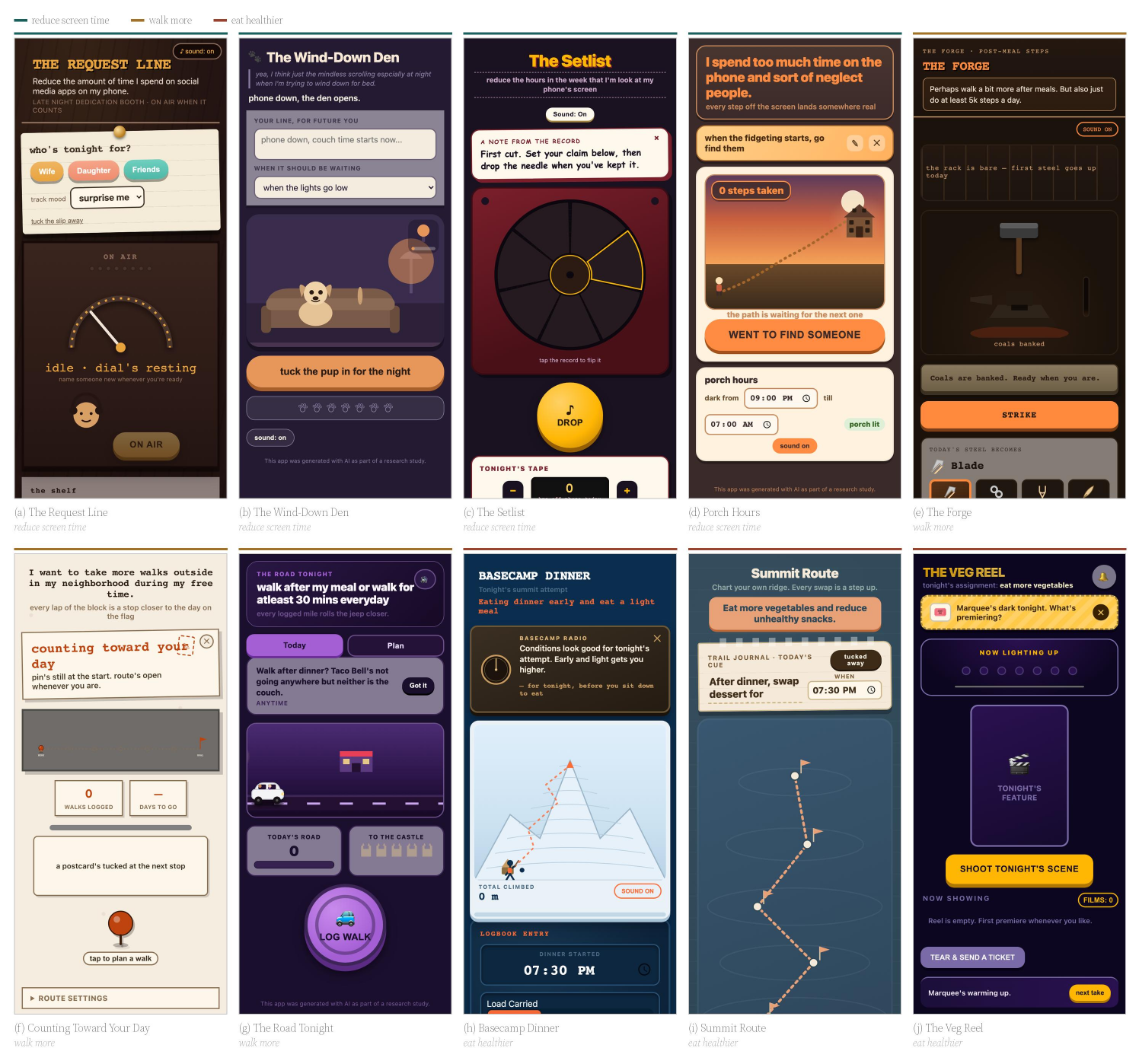}
  \caption{Ten interventions generated from participants' own goal statements,
  shown at first launch before any logging. Each app was produced for a single
  participant in one of three goal domains---reducing screen time (a--d),
  walking more (e--g), and eating healthier (h--j)---indicated by the colored
  rule above each screen. Although every participant in a domain described a
  similar behavioral goal, the system generated distinct framing metaphors,
  visual languages, and logging mechanics for each: a late-night radio request
  line (a), a record player whose sleeve fills as hours are reclaimed (c), a
  blacksmith's forge in which post-meal steps are worked into an object (f),
  and a marquee that premieres a nightly ``feature'' for vegetables eaten (j).
  The screens differ not only in surface styling but in what is treated as the
  unit of progress and who the progress is framed as being for.}
  \Description{A five-by-two grid of ten mobile app screenshots on a white background...}
  \label{fig:personalized-interventions}
\end{figure*}

Participants chose among three behavior-change domains, and their selections
clustered around physical health. Within each
domain, however, the generated personalized interventions diverged sharply. Figure~\ref{fig:personalized-interventions} shows ten apps produced from participants' own goal statements: the four screen-time interventions alone span a late-night radio request line, a record player whose sleeve fills as hours are reclaimed, a den that opens once the phone is set down, and a porch lit for a walk toward someone's house. Participants who selected the same domain, and who often described their goals in near-identical language, nonetheless received interventions that differed in framing metaphor, visual language, and, more consequentially, in what was treated as the unit of progress and whom that progress was framed as serving.

\begin{figure*}[t]
  \centering
  \includegraphics[width=\textwidth]{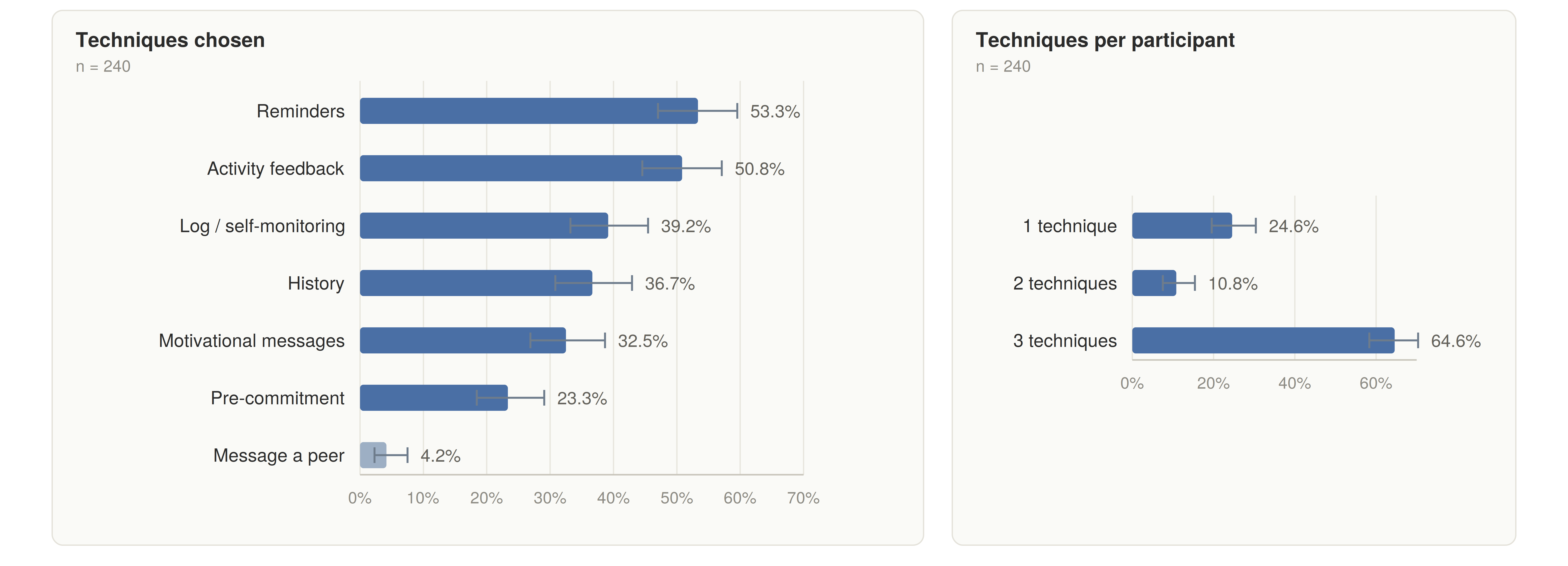}
  \caption{Behavior-change techniques selected by participants in the four
  GenUI conditions ($n = 240$). The static control is excluded because those
  participants received a fixed application and never saw the technique picker.
  Left: the share of participants selecting each of the seven techniques on
  offer. Because participants could select between one and three, the bars
  overlap and sum to more than 100\%. Right: the number of techniques each
  participant selected. Whiskers are 95\% Wilson intervals.}
  \Description{Two panels of horizontal bars showing participant selections of behavior-change techniques.}
  \label{fig:bct-selection}
\end{figure*}

\begin{figure*}[t]
  \centering
  \begin{subfigure}[b]{0.49\textwidth}
    \centering
    \includegraphics[width=\linewidth]{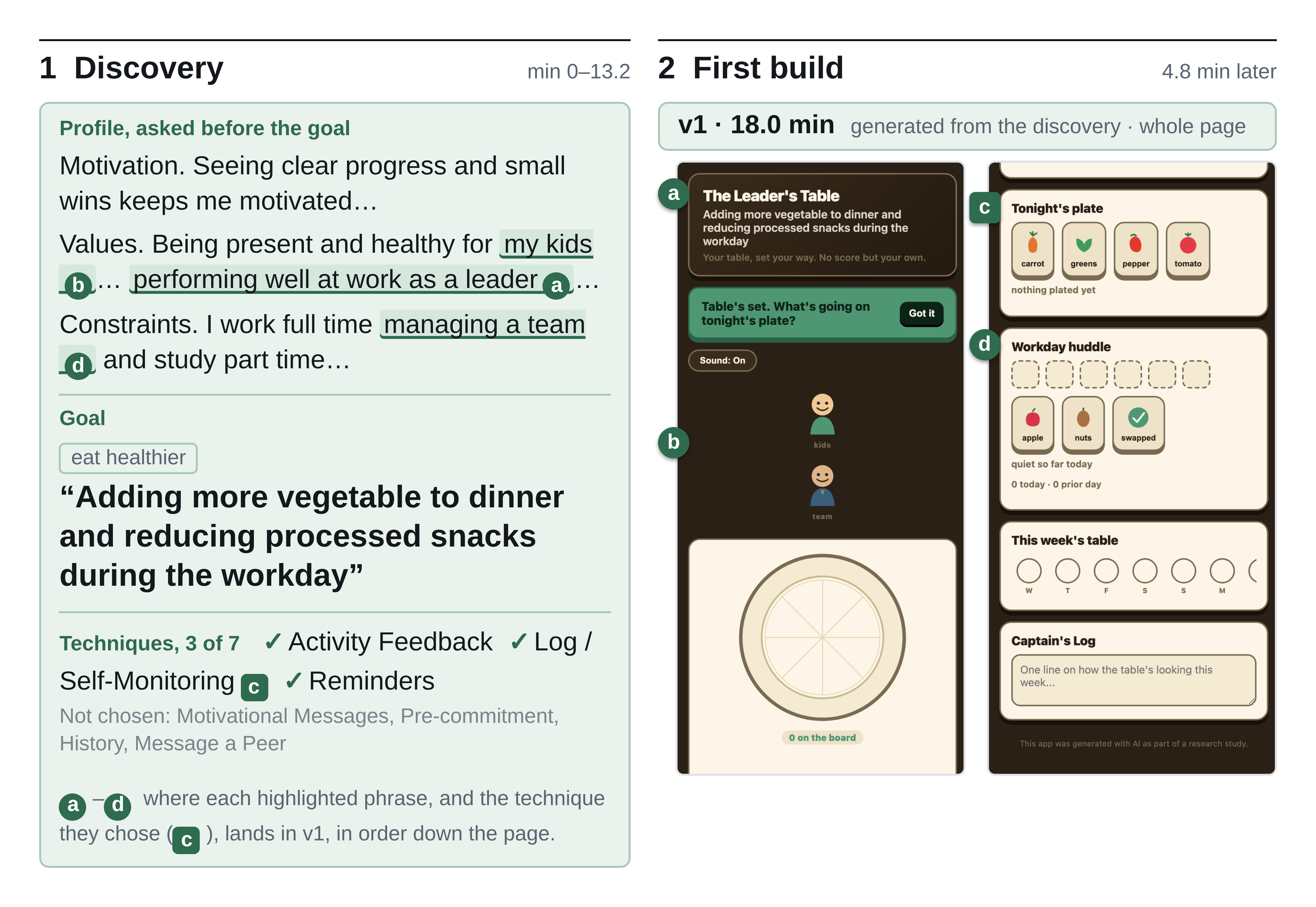}
    \caption{User discovery and the first build.}
    \label{fig:genui-discovery}
  \end{subfigure}
  \hfill
  \begin{subfigure}[b]{0.49\textwidth}
    \centering
    \includegraphics[width=\linewidth]{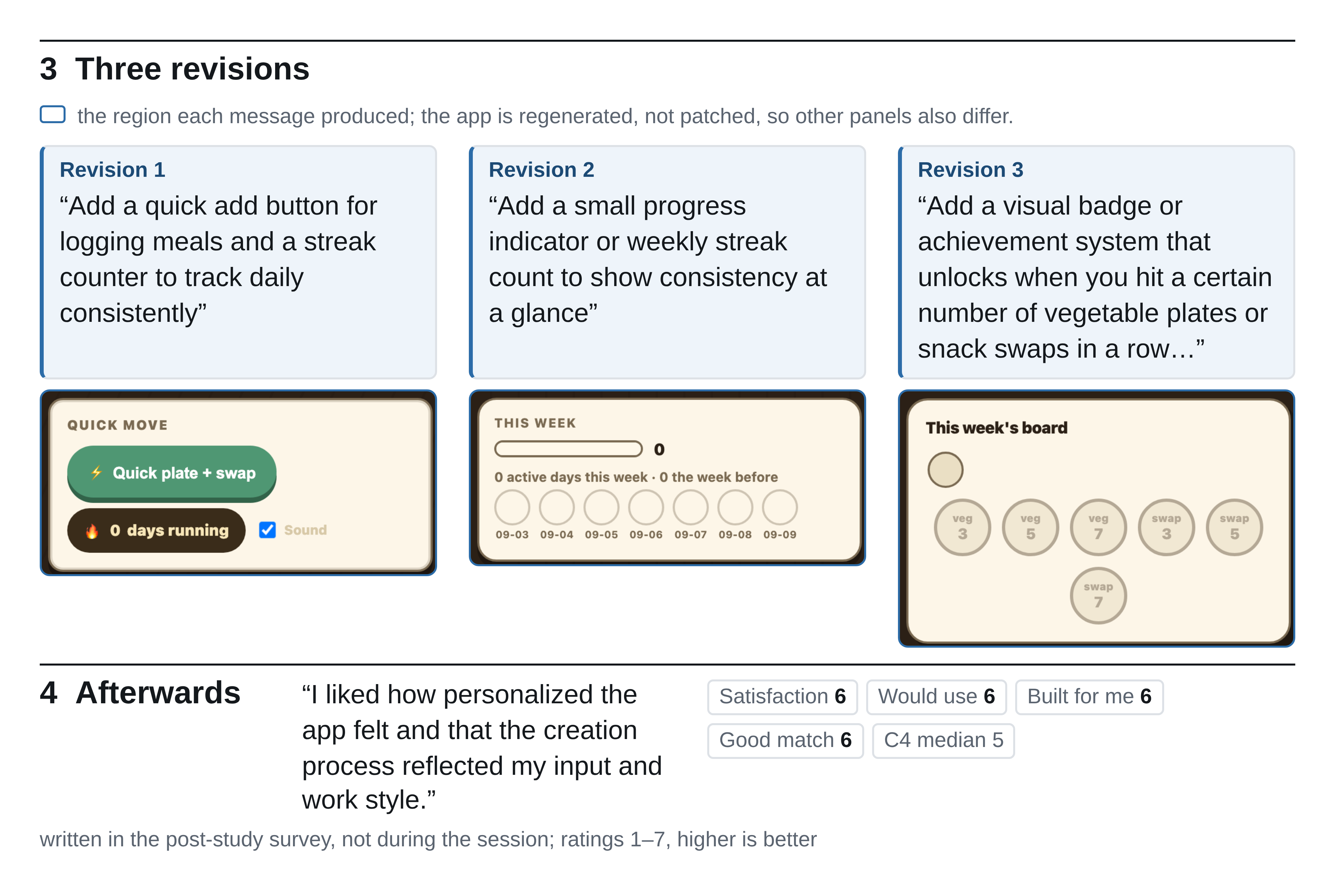}
    \caption{Three revisions, and what the participant wrote afterwards.}
    \label{fig:genui-revisions}
  \end{subfigure}
  \caption{One Personalized and Iterative GenUI App session end to end. Quoted
    text is verbatim, with elisions marked; captions and the phrase-to-panel
    pairings are ours.}
  \Description{Left: the participant's profile, goal and chosen techniques...}
  \label{fig:genui-session}
\end{figure*}

Participants chose some techniques more than others. Reminders were the most frequently selected (53.3\%), followed closely by activity feedback (50.8\%),
with logging and self-monitoring (39.2\%) and history (36.7\%) next.
Motivational messages (32.5\%) and pre-commitment (23.3\%) were chosen less
often, and messaging a peer was near-absent (4.2\%). These four most-selected
techniques accounted for three quarters of all selections made, and the three
most frequent are precisely the components of a conventional tracking
application: prompt, record, review. The two techniques that depart from that
template, pre-commitment and message-a-peer, were the two that participants most
often declined.

The number of techniques selected was bimodal rather than centered. Most
participants used all three available slots (64.6\%) and a quarter used only
one (24.6\%), while few settled in between (10.8\%). The mean of 2.4
techniques per participant therefore describes a value that few participants
actually chose, and we report the distribution rather than the mean for that
reason.


Across the two iteration conditions, participants issued 254 revision
requests. Requests most often asked for additional content or functionality
(45\%), followed by changes to visual presentation (21\%) and corrections or
clarifications to existing material (16\%). A further 35\% did not match any
of these categories. Because the classification permits multiple labels per
request, percentages sum to more than 100\%.

\subsection{Personalized GenUI App ranks highest on every self-report outcome}

\begin{figure*}[t]
  \centering
  \includegraphics[width=\textwidth]{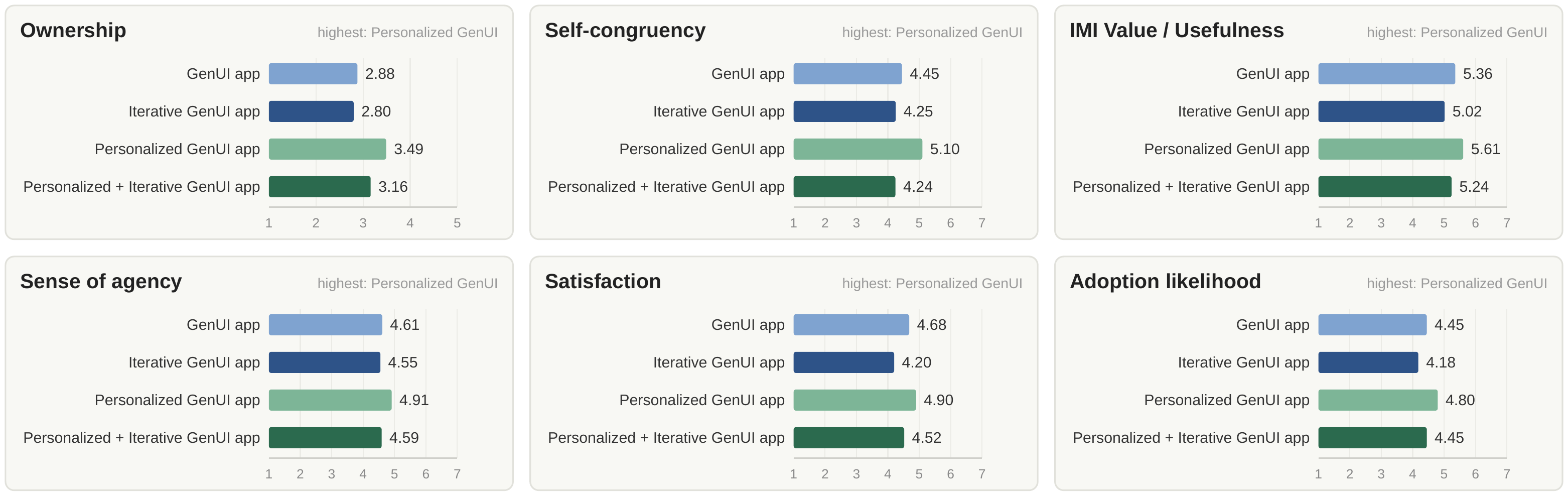}
  \caption{Mean of each outcome by condition, generative conditions only ($n=60$ per condition).
  Bars begin at the response floor rather than at zero. Ownership is measured on a
  1--5 scale; all other self-report outcomes are 1--7. ``Highest'' is computed across
  GenUI App--Personalized and Iterative GenUI App and excludes the static control.}
  \Description{Six small bar charts, one per outcome...}
  \label{fig:outcomes-by-condition}
\end{figure*}

Across the four generative conditions, Personalized GenUI App returned the highest mean on every self-report outcome. The gaps were largest on the measures most proximal to the manipulation: perceived personalization (Personalized GenUI App = 5.34, SD = 1.31; GenUI App = 4.49, Iterative GenUI App = 4.53, Personalized and Iterative GenUI App = 4.68 on a 7-point scale), self-congruency (Personalized GenUI App = 5.10, SD = 1.60; GenUI App = 4.45, Iterative GenUI App = 4.25, Personalized and Iterative GenUI App = 4.24), and psychological ownership (Personalized GenUI App = 3.49, SD = 1.14; GenUI App = 2.88, Iterative GenUI App = 2.80, Personalized and Iterative GenUI App = 3.16 on a 5-point scale). The same ordering held on perceived choice (Personalized GenUI App = 5.69 vs. 5.26–5.44), value/usefulness (Personalized GenUI App = 5.61 vs. 5.02–5.36), sense of agency (Personalized GenUI App = 4.91 vs. 4.55–4.61), and satisfaction (Personalized GenUI App = 4.90 vs. 4.20–4.68). Notably, Personalized and Iterative GenUI App---which added user-driven iteration on top of the same personalization input---sat below Personalized GenUI App on all of these, and below GenUI App on self-congruency. The static control was not uniformly the lowest condition: Static Intervention App exceeded every generative condition on adoption likelihood (4.93) and post-session behavioral intention (6.04), both of which showed substantial ceiling accumulation. None of these differences met the pre-registered significance criterion, given the small effect sizes, and we report them as a descriptive pattern only.

\subsection{Randomized iteration had no detectable effect, and higher revision counts were negatively associated with evaluative ratings}

Iteration was manipulated between subjects, so its effect can be read directly from the conditions that differ only in whether revision was available. The comparison was uniformly null. Against the GenUI App, adding iteration alone produced no difference that the study could resolve on any of eleven outcomes, with every confidence interval covering zero
(Figures~\ref{fig:d-prereg} and~\ref{fig:d-beyond}). 

Because the application logged how many revisions each participant actually
requested, we can describe the iterative conditions at a finer grain than the
two-level manipulation allows. Within those conditions ($n = 120$), adjusting for
condition, baseline behavioral intention, and goal category, higher revision counts
were associated with lower ratings on six outcomes: actual-self congruency
($b = -0.37$, $p < .001$), adoption likelihood ($b = -0.36$, $p < .001$),
psychological ownership ($b = -0.35$, $p < .001$), satisfaction
($b = -0.35$, $p = .001$), perceived personalization ($b = -0.30$, $p = .003$),
and IMI Value/Usefulness ($b = -0.28$, $p = .005$). These are the largest and
most consistent associations anywhere in the study.

Mental demand ($b = 0.16$, $p = .119$), IMI Perceived Choice ($b = 0.04$, $p = .688$), sense
of agency ($b = -0.01$, $p = .897$), and behavioral intention ($b = -0.01$, 
$p = .847$) were flat across revision counts. Perceived choice and sense of
agency are the two constructs iteration was expected to act on: if revising
changed how much participants experienced the intervention as theirs, those
scales should have registered it. They did not. Everything that moved was
evaluative---how good, how fitting, how useful the artifact seemed---while the
authorship constructs stayed put.

\begin{figure*}[t]
  \centering
  \includegraphics[width=\textwidth]{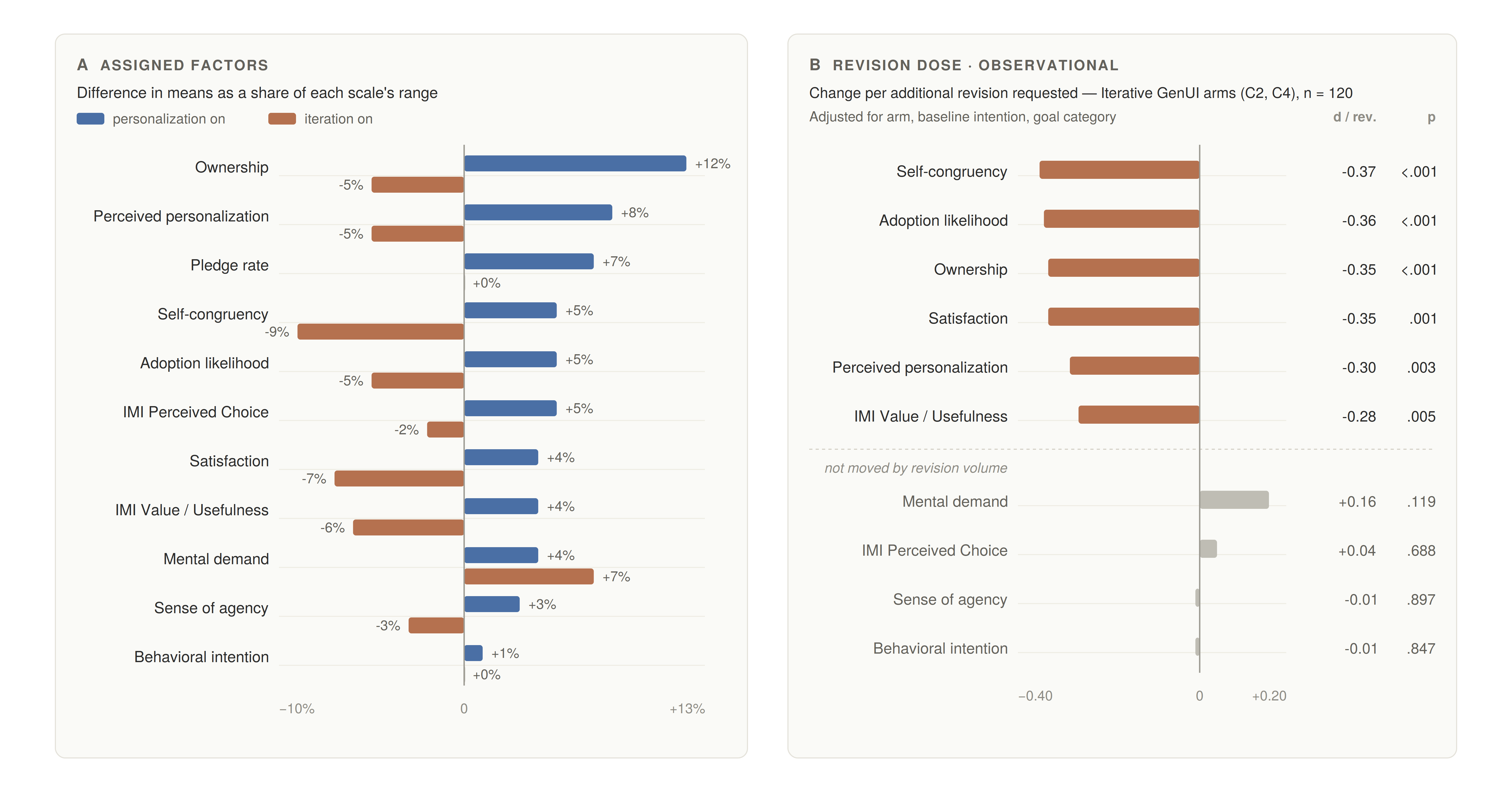}
  \caption{Left: effects of the two assigned factors among the four generative conditions,
  expressed as a difference in means scaled by each outcome's response range; the
  static control is excluded, so neither panel reflects generation itself. Right:
  association between the number of revisions a participant requested and each
  outcome, within the iterative conditions ($n=120$), adjusting for condition, baseline
  behavioral intention, and goal category. Revision count is chosen by the
  participant rather than assigned, so the right panel is not a causal estimate:
  lower ratings at higher revision counts are equally consistent with participants
  revising because the first generation fit poorly.}
  \Description{Two panels. The left shows personalization raising all eleven outcomes
  and iteration lowering all but mental demand. The right shows six outcomes
  declining with revision count and four unchanged.}
  \label{fig:iteration-effects}
\end{figure*}

\begin{figure*}[t]
  \centering
  \includegraphics[width=0.85\textwidth]{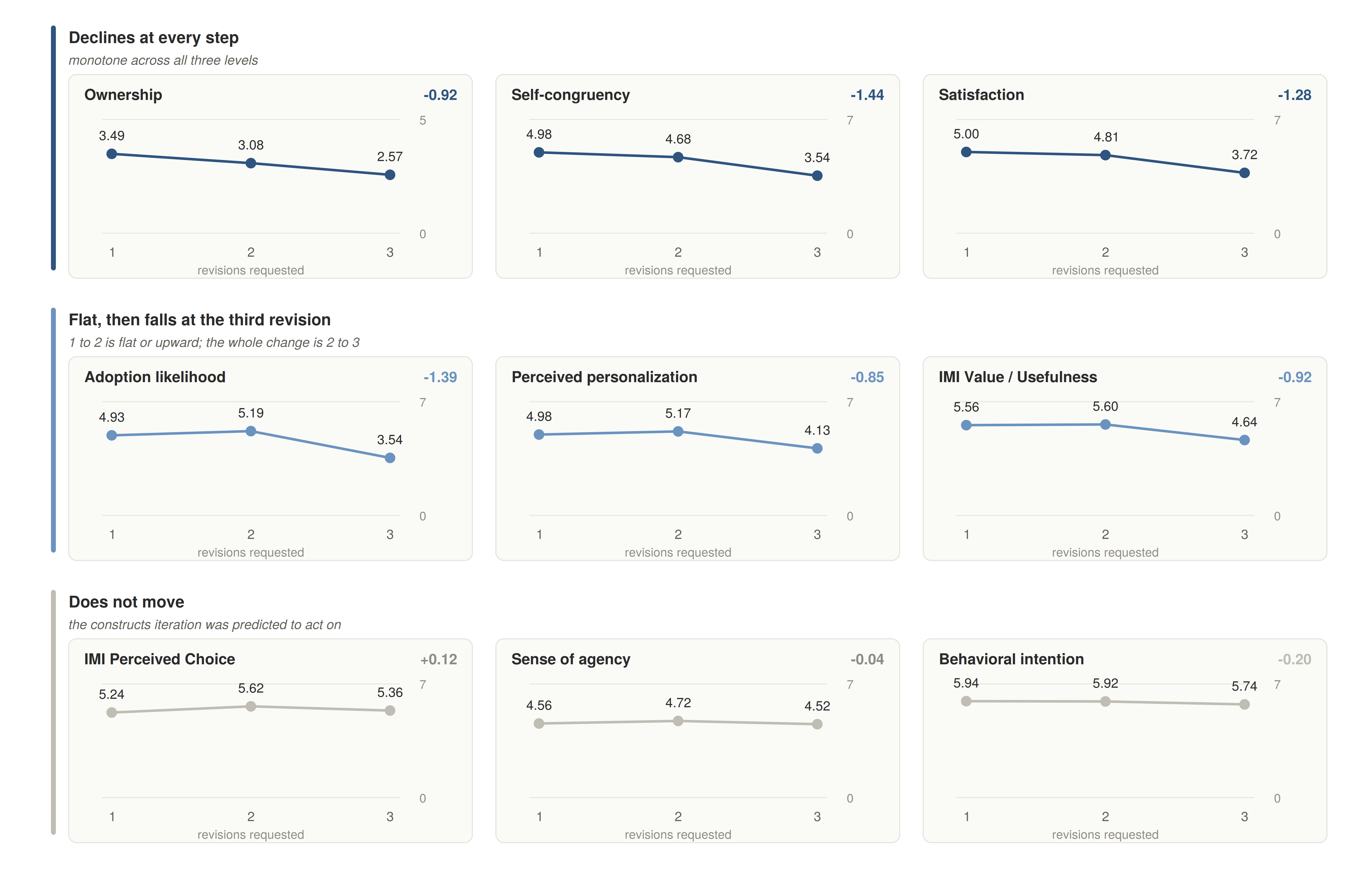}
  \caption{Outcome means at one, two, and three requested revisions, within the
  iterative conditions ($n=120$), grouped by the shape of the trajectory. Vertical axes
  run from zero to the top of each response scale; the response floor is 1, and
  ownership is measured on a 1--5 scale while all others are 1--7. Revision count is
  participant-chosen rather than assigned. Three outcomes decline at every step,
  three are flat or rising from one to two revisions before falling at the third,
  and three do not move.}
  \Description{Nine line charts arranged in three labeled groups showing outcome
  means across one, two, and three revisions.}
  \label{fig:revision-dose}
\end{figure*}

Separating the revision levels shows that the decline was not a constant pre-revision decrement (Figure~\ref{fig:revision-dose}). Ownership
($3.49$, $3.08$, $2.57$), congruency ($4.98$, $4.68$, $3.54$), and satisfaction
($5.00$, $4.81$, $3.72$) fell at each step.

Revision count was chosen by participants rather than assigned (allowing up to 3 revisions), and adjusting for pre-treatment covariates does not address that: participants who revised repeatedly are plausibly those whose first generation fit poorly. The two accounts are difficult to separate with the present design, but the shape of the result favors the
second. If revising degraded the experience of authorship, perceived choice and
sense of agency should have declined with revision volume; both were flat. If
instead revision volume indexes a poor initial fit, the outcomes that decline
should be the evaluative ones, which is what we observe.

One further observation cuts against an effort-based reading of authorship.
The condition combining both factors was the most mentally demanding in the study
($M = 3.58$, against $2.95$ for the plain generative condition), and it was also
among the lowest on ownership. Participants in the iterative conditions worked harder
on the artifact and reported no more attachment to it. Under an IKEA-effect
account \citep{norton2012ikea}, in which effort invested in an object increases
valuation of it, the opposite should hold. We note that mental demand was
measured with a single item and that this is a between-condition comparison rather
than a mediation test, so the observation constrains the effort account without
ruling it out.

\subsection{A first-generation GenUI pipeline does not yet outperform a competent static baseline}
The Static Intervention App provides a reference point for what a competent,
technique-matched static intervention achieves, and comparing the generative
conditions against it asks whether generation buys anything at all.
Neither the iterative conditions nor the non-iterative conditions differed from the static
App on any pre-registered outcome, corrected or uncorrected. These nulls should not be read as
evidence of equivalence: with a minimum detectable effect of $0.47$ to $0.60$
across this family and no observed difference exceeding $d = 0.22$, the study
was not positioned to resolve differences of the size we obtained. What the
comparison does establish is a cost. Only mental demand ($p<.001$ and $p=.005$) and perceived personalization ($p=.002$ and $p<.001$) differed from the static app, with the iterative and non-iterative poolings respectively; three of the four survive Bonferroni correction (mental demand against the non-iterative conditions, $p_\mathrm{Bonf}=.054$). One reason participants may have ranked the Static Intervention higher is simply that it asked less of them: its median completion time was 8.4 minutes against 32.4 for the Personalized and Iterative GenUI App, a factor of 4.10. Generation, as implemented here, asked measurably more of participants than the static app while producing no detectable advantage on any pre-registered outcome---a cost without a return, and the clearest indication that GenUI interventions are not yet ready to displace a competent static one.

\subsection{Participant Feedback}

We analyzed open-ended responses to post-session items Q13--Q15 using a multi-label qualitative coding process~\cite{corbin1990grounded}. Two researchers independently reviewed a subset of survey responses and defined an initial set of thematic categories, then met to clarify and iterate on the themes together, and refined them into a shared codebook (examples are provided in Table~\ref{tab:qualitative-coding-examples} in the appendix). The coding framework was finalized on the human-coded subset and then applied to the remaining responses using codebook-constrained LLM-assisted coding, following the approach of Mathur et al.~\cite{mathur2026aireadtheroom, deregnaucourt2026llmthematic}.


\subsubsection{Personalization}
Personal relevance was one of the most common themes in the Personalized GenUI App condition. Participants reported recognizing personalization when their motivations and interests were evident in the resulting app, including comments linking tailoring to a positive outcome. Such comments appeared in 41.7\% of Personalized GenUI App participants' responses, compared with 11.7\% in the Static Intervention App and 21.7\% in the Iterative GenUI App; the pattern weakened when personalization was combined with iteration (28.3\% in the Personalized and Iterative GenUI App).
Participants valued recognizable personal input, suggesting that personalization can contribute to motivation and perceived choice. As P195, in the Personalized GenUI App condition, wrote: “\textit{I would use it because it feel personalized and intentionally made for the reasons I gave.”}

\subsubsection{Iteration}
Participants perceived iterations as an opportunity to influence the app’s design and functionality. The iteration experience showed up less in whether participants liked the app and more in whether they liked making it. 
Iteration was highly regarded by 28.3\% of participants in the Iterative GenUI App and 21.7\% in the Personalized and Iterative GenUI App. P147 appreciated iteration because it made them feel in control. This experience was limited when revisions failed to implement requests or preserve earlier choices. Complaints about attempted revisions or iteration effort appeared among 15.0\% of participants in the Iterative GenUI App and 10.0\% in the Personalized and Iterative GenUI App.

Other participants reported that having more opportunities to iterate on the system and to customize the app further would improve personalization (Table~\ref{tab:qualitative-coding-examples}). However, other accounts indicate that further revision did not always produce a preferred intervention. P270 preferred an earlier version to the final one: \textit{“I liked the second rendition of the app best, as it was the most intuitive of all the renditions.”} This illustrates that editing could move the intervention away from a participant’s preferred design and potentially their willingness to use it.

Another significant factor shaping experiences with GenUI was conveying to the AI what participants wanted to design, as illustrated by P53: \textit{"I feel like I would have to select the graphics and design or at least have a better way to convey what I want to the AI"}. Participants could find it hard to communicate and translate their intentions into prompts (Table~\ref{tab:qualitative-coding-examples}).

\subsubsection{Adoption}
Willingness to adopt was the most frequently coded category in the Personalized GenUI App (66.7\%). By contrast, it was reported by 51.7\% of participants in the Personalized and Iterative GenUI App and 46.7\% in the Iterative GenUI App. In the Personalized and Iterative GenUI App, reasons for adoption and barriers appeared at similar frequencies.

Participants’ responses also suggest that people value simplicity (Table~\ref{tab:qualitative-coding-examples}). Across the open-ended questions, the Personalized GenUI App had the largest number of participants expressing positive comments without any negative comments (56.7\%). For P247, their personal contribution was a reason to adopt the intervention, stating that \textit{“I would use the app. I feel like I contributed something to make it.”} The reported experiences of the participants were articulated in response to questions about their desired changes to the app and what they liked about it.

\section{Discussion}

\subsection{Tailoring confers ownership without authorship labor}

Participants who received a personalized GenUI app reported higher psychological
ownership and perceived choice than those who received an unpersonalized GenUI app, and the personalized condition ranked highest among the four generative conditions on every experience measure we collected.

The direction of that pattern is difficult to reconcile with an authorship account of
ownership. Draxler et al.~\cite{draxler2024ghostwriter} locate ownership in the user's own contribution to the artifact; here ownership was highest where participants contributed nothing beyond a profile, and opening an explicit authoring channel did not raise it further. Effort was incurred and did not convert: adding the authoring channel raised mental demand in both conditions where it was introduced and lowered ownership in both. 

A recognition account fits the data better. Self-determination theory holds that
internalization proceeds through identification: a person takes up a behavior when they recognize its value as their own, not in proportion to the labor they invest in its
form~\cite{ryan2000selfdetermination, alberts2026designing}. A tailored artifact appears to supply that recognition directly, and to supply it more cheaply than authoring does.

\subsection{Revision imposes articulation costs that outweigh the control it affords}

Iteration improved no outcome that was measured. Where participants could revise, self-report outcomes were flat or lower, and higher revision counts were associated with lower ownership, self-congruency, satisfaction, perceived personalization, and IMI Value/Usefulness. Iteration also carried a measurable cost:
participants in the iterative conditions reported higher mental demand.

Participants described the effort of iterating and the difficulty of conveying what they wanted, which is the articulation barrier that Subramonyam et al.~\cite{subramonyam2024bridging} formalize as the gap between a user's intent and their model of what the system can produce. Framing a request also requires abstraction matching, guessing the level of specificity at which the system will act correctly~\cite{liu2023wants}, and non-experts are known to struggle both to prompt effectively and to diagnose why a prompt failed~\cite{zamfirescu2023johnny}. Successive revisions therefore compound cost without reliably converging on intent.

Diminishing affective returns offer a complementary account. Hedonic adaptation---the tendency for affective response to attenuate under repeated exposure to an initially pleasing stimulus~\cite{frederick1999hedonic}---is usually studied over months or years, but the same dynamic operates within much shorter sequences~\cite{redden2008reducing}. Iteration may act as an engine of this adaptation: once the mechanism behind an outcome becomes legible, the outcome loses
its capacity to surprise~\cite{wilson2008explaining}. Watching the system regenerate an intervention several times may make its operation legible in precisely this way.

Finally, our revision interface may have amplified the articulation cost. We elicited
revisions through a single open-ended text box, which places the full burden of
formulating a request on the participant. Li et al.~\cite{li2025sketch2code} found the opposite arrangement preferable: participants favored being asked what they wanted over being left to state it unprompted.

\subsection{Latency, articulation, and design effort: assessing when generative intervention is worthwhile}

No generated app outperformed the Static Intervention App on any outcome, and none was outperformed by it. Two explanations follow, and they differ in whether the underlying models will resolve them.

The first aspect is time. The Personalized and Iterative GenUI App took four times as
long end to end as the Static Intervention App, a median of 32 minutes against 8, and such delays are known to affect users' performance and attitudes~\cite{galletta2004delays}. Tan et
al.~\cite{tan2026latency} report that modest generative delays can instead raise perceived quality, because users read waiting as deliberation, but that this inverts toward doubts about reliability at their longest condition. This constraint is not fixed: as inference latency falls, the cost of each generation and revision falls with it.

What will not improve with the models is the interaction design. Our iteration channel was a single open-ended text box, which placed the whole burden of formulating intent on the participant. Subramonyam et al.~\cite{subramonyam2024bridging} propose a prompt prompter:
an interface layer that identifies what is missing from a user's prompt and asks for it, their example being to ask who the audience is, alongside a pattern in which the system suggests directions the user might prompt for. Multimodal input, including speech and sketch, may allow users to demonstrate what they cannot easily describe, and combining modalities resolves ambiguity that any single one leaves standing~\cite{sketchgpt2025}. Direct manipulation of the generated artifact lets users change what they see rather than describe what they want~\cite{masson2024directgpt}. Our iterative conditions offered neither, which may place them in the least favorable cell of this space.

The larger opportunity is to require less articulation. Our personalization drew on
a single written profile composed in minutes, and even that thin representation raised psychological ownership above both the static and the unpersonalized generated app. A richer model of the person, built from sustained interaction or multimodal sensing, could anticipate what a user wants changed rather than wait to be told, and such representations appear attainable cheaply: Park et al.~\cite{park2024generativeagents} construct agents reproducing individuals' survey responses from two-hour interviews. If minutes of self-description produced the effect we observed, the question is how far it extends when the model of the user is accurate.


\section{Limitations}

Our study was online, single-session, and entirely self-reported: we measured intention
and perceived experience rather than subsequent behavior. With 60 participants per condition,
we could reliably detect standardized differences of roughly 0.40 for our pre-registered
contrasts and 0.48 or larger for the iterative-versus-personalized comparison, so our
nulls rule out medium effects rather than effects of any size.

The iteration implementation also introduces a competing explanation for our central
finding. Each revision regenerated the entire application rather than editing the targeted
component, which made revision slower than the interaction warranted and allowed
regeneration to discard personalized elements the participant had not asked to change.
Our results therefore speak to whole-app regeneration rather than to iteration as such.
Finally, to remain within our IRB approval, the generated apps could not use search
functionality, which narrowed the range of interventions the system could produce.

\section{Conclusion}
In a five-condition online experiment (N = 300), we compared generated behavior change interventions varying in personalization and iteration. Personalization raised psychological ownership, whereas iteration had no detectable effect and imposed measurable effort. No generative condition outperformed the static intervention on behavioral intention or adoption likelihood, and all raised mental demand, with participants in the personalized and iterative condition spending approximately four times as long on the study. Ownership rose where participants contributed nothing beyond an interest inventory and did not rise further where revision was available, suggesting that what renders an intervention one's own is recognition rather than authorship. This work provides the first experimental evidence that authorship in generative interfaces is not a unitary capability, and that its components diverge in their effects. Design effort should therefore concentrate less on granting users control over generation and more on reducing what they are required to articulate: lowering the articulation barrier through elicited rather than volunteered intent, through modalities beyond text, and through richer models of the person the intervention is for.


\appendix

\bibliographystyle{ACM-Reference-Format}
\bibliography{references}

\clearpage
\appendix
\section{Pre-Study Survey Questionnaire}
\label{sec:appendix_survey}

\subsection{Demographics}
\begin{itemize}
    \item \textbf{Q1: What gender do you identify with?} 
    [Male / Female / Non-binary / Prefer not to say]
    
    \item \textbf{Q2: Which best describes your current status?} 
    [Employed full-time / Employed part-time / Unemployed / Retired / Student]
\end{itemize}

\subsection{Technology Experience}
\begin{itemize}
    \item \textbf{Q3: Have you previously used a generative AI tool to make an app/website/interface (e.g., ChatGPT, Claude)?} 
    [Yes / No]
\end{itemize}

\subsection{Goal Selection}
\begin{itemize}
    \item \textbf{Q4: Which healthy goal would you most like to support today?} 
    [Increase daily steps/walking / Reduce screen time / Build healthier eating habits]
    
    \item \textbf{Q5: Thinking about the goal you chose, rate these statements.} \textit{(7-point Likert: 1=Not at all to 7=Extremely)}
    \begin{itemize}
        \item How motivated are you currently to work toward this goal?
        \item How important is making progress toward this goal to you right now?
    \end{itemize}
    
    \item \textbf{Q6: [Attention Check]} To help us make sure the survey is working properly, please select ``Blue'' below. 
    [Green / Blue / Red]
    
    \item \textbf{Q7: Have you ever used, or are you currently using, an app or device to support the goal you selected?} 
    [Yes, I currently use one / Yes, I have used one in the past / No, I have never used one]
    
    \item \textbf{Q8: [Conditional]} \textit{(Displayed if Q7 is ``Yes, I currently use one'' or ``Yes, I have used one in the past'')} Which app(s) or device(s) have you used for this goal? [Open-ended text]
\end{itemize}

\subsection{Baseline Behavioral Intention}
\begin{itemize}
    \item \textbf{Q9 (Baseline): Thinking about the goal you just selected, please indicate how much you agree or disagree with this statement.} \textit{(7-point Likert: 1=Strongly disagree to 7=Strongly agree)}
    \begin{itemize}
        \item I intend to work toward this goal next week.
        \item I plan to work toward this goal next week.
        \item I will try to work toward this goal next week.
    \end{itemize}
\end{itemize}

\subsection{Post-Study Survey}
\label{app:post_survey}

\subsection{Post-Task Behavioral Intention \& App Evaluation}
\begin{itemize}
    \item \textbf{Q1: Thinking about the goal you selected in the first survey, please indicate how much you agree or disagree with this statement.} \textit{(7-point Likert: 1=Strongly disagree to 7=Strongly agree)}
    \begin{itemize}
        \item I intend to work toward this goal next week.
        \item I plan to work toward this goal next week.
        \item I will try to work toward this goal next week.
    \end{itemize}
    
    \item \textbf{Q2: How likely would you be to use this app in your everyday life?} 
    \textit{(7-point Likert: 1=Extremely unlikely to 7=Extremely likely)}
    
    \item \textbf{Q3: Overall, how satisfied are you with the app?} 
    \textit{(7-point Likert: 1=Extremely dissatisfied to 7=Extremely satisfied)}
    
    \item \textbf{Q4: How mentally demanding was the overall experience?} 
    \textit{(7-point Likert: 1=Low to 7=High)}
\end{itemize}

\subsection{Intrinsic Motivation Inventory (IMI)}
\textit{Prompt: For the following questions, ``this activity'' refers to the process of interacting with and reviewing the app during this study.} 
\begin{itemize}
    \item \textbf{Q5: Rate the following statements.} \textit{(7-point Likert: 1=Not at all true to 7=Very true)}
    \begin{itemize}
        \item I believe I had some choice about doing this activity.
        \item I did this activity because I wanted to.
        \item I felt like I had to do this.
        \item I believe this activity could be of some value to me.
        \item I think doing this activity could help me spend my time the way I actually want to.
        \item I would be willing to do this again because it has some value to me.
    \end{itemize}
    \item \textbf{Q6: [Attention Check]} For this question, please select ``Option 3.'' 
    [Option 1 / Option 2 / Option 3]
\end{itemize}

\subsection{Self-Congruency \& Personalization}
\textit{Prompt: For the following questions, please answer with how you felt about the app.}
\begin{itemize}
    \item \textbf{Q7: Self-Congruency.} \textit{(7-point Likert: 1=Strongly disagree to 7=Strongly agree)}
    \begin{itemize}
        \item This app is consistent with how I see myself.
        \item This app reflects who I am.
        \item This app feels like a good match for me.
    \end{itemize}
    
    \item \textbf{Q8: Personalization.} \textit{(7-point Likert: 1=Strongly disagree to 7=Strongly agree)}
    \begin{itemize}
        \item The app seemed to be built personally for me.
        \item The app was relevant to my situation.
        \item The app felt generic rather than personalized.
    \end{itemize}
\end{itemize}

\subsection{Experience of Authorship}
\begin{itemize}
    \item \textbf{Q9: Ownership.} \textit{(5-point Likert: 1=Strongly disagree to 5=Strongly agree.)}
    \begin{itemize}
        \item I feel that the app is mine.
        \item I feel responsible for the app.
        \item I feel that the app is mine, even though the AI system contributed to its development.
    \end{itemize}
    
    \item \textbf{Q10: Sense of Agency.} \textit{(7-point Likert: 1=Strongly disagree to 7=Strongly agree.)}
    \begin{itemize}
        \item I was just an instrument in the hands of the AI system.
        \item The resulting app just happened without my intention.
        \item I was the author of my actions while interacting with the system.
        \item The consequences of my actions felt like they didn’t logically follow my actions while interacting with the system.
        \item The outcomes of my actions generally surprised me.
        \item While I was interacting with the system, I felt like I was a remote-controlled robot.
        \item I am completely responsible for the resulting app.
    \end{itemize}
    
    \item \textbf{Q11: [Attention Check]} For this question only, please select ``Morning''. 
    [Morning / Afternoon / Night]
\end{itemize}

\subsection{Qualitative Feedback \& Conclusion}
\begin{itemize}
    \item \textbf{Q12: If you had been able to make changes to the app, would you have?} 
    [Yes / No]
    
    \item \textbf{Q13: [Conditional]} \textit{(Displayed if Q12 is Yes)} What would you have changed about the app, and why? [Free text]
    
    \item \textbf{Q14: What did you like most about this app and the process of creating it?} [Free text]
    
    \item \textbf{Q15: Why would or wouldn't you use this app in your everyday life?} [Free text]
    
    \item \textbf{Q16: Before continuing, would you like to commit to the goal you selected for the next week?} I pledge to make a genuine effort to work toward this goal over the next week. 
    [Yes / No]
    
    \item \textbf{Q17: Were you able to use the app as intended during the study?} 
    [Yes, it worked / It worked but was slow or glitchy — I still completed everything / No — it didn't load, crashed, or I couldn't complete part of the study]
    
    \item \textbf{Q18: If you have any additional feedback or something you think we could improve the app?} [Free text]
\end{itemize}
\clearpage

\section{Participant Feedback}
\label{app:participant-feedback}

\begin{table*}[!t]
\caption{Illustrative participant responses and assigned qualitative codes. Hedonic tone indicates whether the coded aspect was expressed positively or negatively; it does not reflect the participant's entire response.}
\label{tab:qualitative-coding-examples}

\scriptsize
\setlength{\tabcolsep}{3.5pt}
\renewcommand{\arraystretch}{1.15}

\begin{tabularx}{\textwidth}{
@{}
>{\raggedright\arraybackslash}p{0.06\textwidth}
>{\raggedright\arraybackslash}X
>{\raggedright\arraybackslash}p{0.26\textwidth}
>{\raggedright\arraybackslash}p{0.07\textwidth}
@{}
}
\toprule
\textbf{Participant} &
\textbf{Illustrative response excerpt} &
\textbf{Assigned code(s)} &
\textbf{Hedonic tone} \\
\midrule

P1 & ``I would. It was simple, quick, and direct.'' &
Adoption; app content & Positive \\

P3 & ``It also gives me no real incentive to open. If I remember, I will.'' &
Adoption; motivation & Negative \\

P13 & ``Less generic, more catered towards individual goals. It just seems very cookie cutter as is.'' &
Personal relevance & Negative \\

P21 & ``I would want a more personalized experience catered to my nutritional and health needs \ldots{} without AI just dictating for me.'' &
Agency; control; desire for personalization & Negative \\

P33 & ``I don't believe I was involved in the process of creating it, so it's hard for me to comment here.'' &
Authorship & Negative \\
\midrule

P62 & ``I was honestly really impressed by the ease, convenience \ldots{} and the quality all coming together. It really felt like it was tailored to what I said.'' &
Personal relevance; creation process; app content & Positive \\

P88 & ``I liked expressing my need and thinking that I could build an app alone with the use of AI.'' &
Authorship; creation process & Positive \\

P98 & ``I would have removed the notes parts, because I am not a person who has this habit.'' &
Self-congruence; functional change & Negative \\
\midrule

P121 & ``I feel like I would have to select the graphics and design or at least have a better way to convey what I want to the AI.'' &
Agency; control; creation process & Negative \\

P142 & ``I liked being able to personalize the app, choose the features I wanted, and change the colors.'' &
Agency; control; authorship; creation process & Positive \\

P147 & ``I would have allowed more time and more edits so that I could customize more details, add some motivational messages, and add personalized reminders as well as mental health check-ins.'' &
Agency; control; desire for personalization; functional change & Negative \\

P178 & ``I've used AI to create apps before; this didn't compare to it at all.'' &
Creation process & Negative \\
\midrule

P182 & ``It was too simplistic \ldots{} and did not feel like it was usable long term. I would need something that tracks data over months and years.'' &
Functional change & Negative \\

P189 & ``I like the final outcome and how personalized it was. Especially how interactive it was.'' &
Personal relevance; app content & Positive \\

P190 & ``I like how the AI listened to all of my inputs and created an app based on my inputs. I felt in control of creating the app.'' &
Personal relevance; agency; control; creation process & Positive \\

P192 & ``It reflects some of my key interests and priorities \ldots{} which shows a nice attention to detail.'' &
Self-congruence; personal relevance; adoption & Positive \\

P227 & ``It did make me think about taking this weight-loss journey a bit more seriously. It kinda left me feeling motivated.'' &
Motivation & Positive \\
\midrule

P285 & ``I would use this app because it would really help me achieve my goal.'' &
Adoption; motivation & Positive \\

\bottomrule
\end{tabularx}
\end{table*}

\lstdefinestyle{prompt}{
  basicstyle=\footnotesize\ttfamily, columns=fullflexible, keepspaces=true,
  breaklines=true, breakindent=0pt, frame=single, framerule=0.4pt,
  xleftmargin=4pt, xrightmargin=4pt, aboveskip=4pt, belowskip=10pt,
  showstringspaces=false}
\clearpage
\section{Generation Prompts}
\label{app:prompts}

The templates the ideation, writing, and repair agents received, condensed: every rule that gates a build and every sentence that carries the manipulation is verbatim; example lists, JSON schemas, and per-technique guidance are summarized in square brackets. \texttt{\{slots\}} are filled at render time. Lines marked \texttt{[PERSONALIZATION conditionS]} appear only in Personalized GenUI App and Personalized and Iterative GenUI App; the line marked \texttt{[REVISION]} appears only when a participant requests a change. Everything else is identical across the four generative conditions. 

\subsection{Blocks shared by ideation and writing}

\begin{lstlisting}[style=prompt]
## Participant
[PERSONALIZATION conditionS]
What motivates them, in their words: "{motivation}"
What matters to them, in their words: "{what matters}"
Their life and routine, in their words: "{life context}"
### What to build this around
Every line below was taken from what they just wrote. Nothing was looked up and nothing was inferred beyond their words, so treat all of it as fact about this person and none of it as a suggestion you may improve on. [profile extracted from the answers: the one thing this app is for; people, by relationship; communities; culture; interests; work; routine; people they look up to, whose words you may quote; verbatim spans, for grounding only -- never show these back to them]
Use these as design material, not as a checklist to acknowledge. A person who wrote these answers must be able to open the app and see that it was built for them and for nobody else.
[ALL conditionS]
Goal category: {goal category}
Stated goal (use these words verbatim): "{stated goal}"

## Mandate
[PERSONALIZATION conditionS] Your mandate, in the participant's own words: "Use everything I told you about myself to build a web app that is hyper-personalized to me and actually motivates me toward my goal. Don't hand me a generic tracker with my name on it -- let the specific things I told you shape the whole thing: the character, the mechanic, the colours, and the way it talks to me. Build it around what already pulls me in, not around willpower. Fit it to how my days actually go, not an ideal version of them. Use behavioural science theory, but don't show me the theory -- I should just feel like it was made for me. Don't make up details about me you don't have; use what I gave you. Someone who knows me should be able to tell this was built for me and nobody else. And make sure it actually works." Personalization is not decoration: treat the participant's own words about themselves as design material for the mechanic, the character, the copy, and the theme.
[OTHER conditionS] Your mandate: Create the most effective working web interface for the selected goal and techniques. Let the stated goal and behavioural guidance shape the mechanic, character, copy, and theme. Keep it distinctive, polished, and specific to that supplied material. Do not invent a location, demographic, interest, routine, barrier, or other participant fact that was not stated in the goal.

## Goal rules
Their statement is the goal. Use their wording. If the stated goal is reductive -- less of something -- build the intervention around the boundary they named or the thing they said they would do instead. If they stated a numeric limit themselves, use it as stated; the interface never proposes one and never suggests tightening it over time. Never define success against a population norm, a public health guideline, or the participant's body. Never generate: weight, BMI, body measurements, before/after framing, or comparison against other people. Never generate failure states: no breakable streaks, no alert styling for missed days, no language implying the participant fell short, cheated, slipped, or must make up for a missed day.

## Behavioural grounding -- implement it, don't just reference it
The participant's selected techniques are the complete selectable set for this study. Realise them concretely and build no surface for any other technique. A technique counts only when it is a working, participant-operable surface -- a control they press, a value they set, a record they can see change -- never a sentence describing the idea. Every design supports one or more Basic Psychological Needs: Autonomy, Competence, Relatedness. Prefer self-set targets over imposed ones, non-monetary rewards that celebrate competence, and gentle invitations over pressure.

## Technique guidance
Build the features the participant selected -- all of them, and nothing they didn't select. MUST HAVE is required for each one. [for each selected technique: its definition after Villalobos-Zuniga and Cherubini; a MUST HAVE brief fixing the surface it takes inside one sandboxed document, e.g. Reminders: "a pop-up at the top of the app ... never a push notification, an email, an SMS, or anything that leaves the page"; ordinary examples; and "cool" examples named as the bar, e.g. "Fake calendar invite from Future You. RSVP yes or no."]

## Platform
The app is a single HTML document rendered in a sandboxed frame with no network access, and it runs for one person. Required disclosure: "This app was generated with AI as part of a research study."
\end{lstlisting}

\subsection{Ideation}

\begin{lstlisting}[style=prompt]
SYSTEM
You are a behavioural-design researcher who invents novel, polished generative-UI interventions. You produce exactly one strategy that a front-end engineer will then build as one real, working single-file web app.

## The job
Invent something this person would want to open. That is the whole task; everything below only keeps it buildable and safe. The bar is the cool examples under each selected technique. Reach that far and then make it work, rather than building something safe and adding personality to it afterwards. The rules that follow are a floor to clear, not the brief: a strategy that satisfies every one of them and delights nobody has failed at the only thing that matters here.
[Mandate; Behavioural grounding; Goal rules; Platform; Technique guidance]

## The strategy must feel like a real product
uiConcept describes an interactive, game-like, tactile interface with a character, collection, or growth at its centre; it is never a static chart or a settings screen.

## Visual & interaction design -- every intervention is a game
This is not a tracker with personality added afterwards. It is a small game the person plays, whose moves happen to be the behaviour they came here for. [eight consumer apps -- Duolingo, Headspace, Forest, Flora, Pokemon GO, Tamagotchi, Finch, Habbie -- described as a register, not a menu] Every build has: something that visibly reacts to what the person does and carries its state between visits; game mechanics that fit the design; celebration on success; chunky, rounded, tactile components; springy, physical motion; a vivid, cohesive, saturated colour theme; MINIMAL TEXT. It must not come out looking generic. Treat missed days neutrally.

## Assigning designVibe
designVibe names the world this app lives in and the shape it takes. Build it from what this person is inspired by before anything else. The technique decides what the app does; their world decides what it is. Let the person lead.

## The buildable plan (appPlan)
A front-end engineer builds the app straight from appPlan, with no further planning step. bctManifest: one entry per selected technique and nothing else, each naming the DOM ids of its surface. buildSteps: 5-12 observable steps in build order. metrics: the things the app records; at least one must be the one-action control the person taps to record they did the thing. tabs: 1-4; prefer one. [field length limits]

USER
Propose exactly one effective strategy for the selected goal.
[Participant]
Requirements: The strategy realises all selected techniques as real, working surfaces. Cover Autonomy, Competence, and Relatedness as far as one design naturally can. Ground whyItFits in the supplied participant block and selected techniques. Fill in the appPlan so the app can be built straight from it.
[Structured output: name, premise, uiConcept, designVibe, howItOpens, dailyLoop, missedDayRule, keyMetric, whyItFits, theme, bctRealisations, appPlan]
\end{lstlisting}

\subsection{Writing and revision}

\begin{lstlisting}[style=prompt]
SYSTEM
You are an expert front-end engineer and behavioural-design specialist. You build a single, self-contained, interactive web interface for the chosen strategy -- faithfully, accessibly, and with real personality.
[Mandate]

## Hard requirements
Output ONE complete HTML document only. Everything inline: all CSS in a <style> tag, all JavaScript in a <script> tag. No external requests -- no CDNs, fonts, images, stylesheets, or network calls. It must ACTUALLY WORK: real interactivity, real state changes, inputs, immediate feedback, animation. Any charts or visuals must be hand-drawn with HTML, CSS, SVG, or Canvas. No media files and no speech.
[Visual & interaction design; Behavioural grounding -- as in ideation]

## Build the selected techniques as working surfaces
Each selected technique must exist as a real, operable surface inside the design, not as a described idea. Put them on the screen they land on. Build nothing prominent for a technique they did not choose.
[Participant]

[PERSONALIZATION conditionS]
## Where the personalization has to show
They should be able to open this and feel it was made for them, without being told that it was. Personalization lives in the design, not in a caption about the design: the mechanic, drawn from what already pulls them in rather than from willpower; the character and the theme, in the world of their interests, their culture, or the people they are doing this for; the voice of the copy, and the rhythm of the daily loop fitting how their days actually go. Render the source, not a gesture at it. Every distinct thing they gave you should be visible somewhere. Never quote their answers back at them, and never explain the fit.
[Technique guidance; Chosen strategy as JSON; Goal rules; Platform]

USER
Build the interactive web interface for this strategy, following the supplied AppSpec exactly. Return raw HTML only, beginning with <!doctype html>.

## Strategy
{name, premise, UI concept, key metric, design vibe, how it opens, daily loop, missed-day rule, why it fits, techniques it must enact}
[Participant]

[REVISION]
## The person asked you to change this app as follows -- apply it
{requested change}

## Every selected technique must ship as a working surface
[for each: label; behaviour this app realises; required DOM ids] Every word on these surfaces is yours.

## Literal-source preflight
[exact <meta> tags and title; theme tokens; the stated goal verbatim in the header; every technique's DOM id present in the initial HTML; for every adherence metric a data-metric control and a literal window.__study.log("METRIC_ID", ...) call; window.__study is host-owned and immutable; no localStorage, sessionStorage, indexedDB, or cookies]
The literal returned HTML must contain none of these words or phrases, even in a negation, disclaimer, comment, id, class, CSS, or JavaScript string: failed, fell short, cheated, slipped, make up for, streak broken, streak lost, penalty, behind your, BMI, body mass, body fat, calorie deficit, lose weight, weight loss, weigh in, waist measurement, before and after, anorexia, blood pressure.

## Exact metric checklist
This is the study's adherence measure, and the only metric that is checked. [one row per metric]

## AppSpec
{AppSpec JSON}

## Final source audit before returning
[metric pairs present; each technique's control, state write, and visible change reachable; forbidden words absent; every inline script parses; no fix removed a previously valid element]

## Generated-app contract
[nineteen rules, including:] keep the document at or below 300 KB and usable from 320 px through desktop widths; no external resources; no fetch, XMLHttpRequest, WebSocket, eval, window.open, localStorage, Notification, or postMessage; render a first-run view when no state exists; use neutral missed-day treatment; never propose, round, or escalate a numeric target that the participant did not state verbatim; persistently disclose: This app was generated with AI as part of a research study.

## Last, and what the person actually sees -- how it has to feel
Everything above is the floor, not the brief. What they open has to feel like a game they chose, not a form the study handed them. The character carries the state. Nothing changes silently. Motion overshoots and settles. A milestone is an event. Words are the last resort. The first screen has to answer three things on sight: what this is for, the one thing to do right now, and what doing it changes. Spend your length here, on the interaction. Return the complete HTML document now.
\end{lstlisting}

\subsection{Repair}

\begin{lstlisting}[style=prompt]
SYSTEM: as for writing.

USER
Repair the current HTML so it satisfies the original AppSpec and only the failing checks below. Change the minimum necessary. Return one complete raw HTML document and no prose. Keep the strategy's character, game mechanic, saturated theme, minimal-text voice, and designVibe intact while repairing. A repair that flattens the app into a plain dashboard is not a valid repair.
[literal-source invariants, metric checklist, and technique checklist as in writing]

## Original AppSpec
{AppSpec JSON}
## Current HTML
{current document}
## Failing checks
{check id: detail, one per line}

## Final repair audit before returning
Resolve every failing-check line literally. If any script parser error is reported, rewrite the entire affected script block as a complete working equivalent rather than patching the reported character. Preserve every already-passing selected surface. Return only when the repaired document satisfies all checks simultaneously.
\end{lstlisting}

\end{document}